\documentclass[aps,pre,preprint,groupedaddress,showpacs,10pt]{revtex4}
\usepackage{amsmath,epsfig,latexsym,graphics,subfigure,makeidx}
 \usepackage[dvips]{graphicx}
\newcommand{\be}{\begin{equation}}
\newcommand{\ee}{\end{equation}}
\newcommand{\bea}{\begin{eqnarray}}
\newcommand{\eea}{\end{eqnarray}}
\newcommand{\noi}{\noindent}

\begin{document}

 

\title{\bf{Coarsening dynamics of ternary amphiphilic fluids and the 
self-assembly of the gyroid and sponge mesophases: lattice-Boltzmann 
simulations}}

\author{N\'elido Gonz\'alez-Segredo
\footnote{\small{\texttt{n.gonzalez-segredo@ucl.ac.uk}. Also at
Departament de F\'\i{}sica, Universitat Aut\`onoma de Barcelona,
08193 \mbox{Bellaterra}, Barcelona, Spain.}\normalsize}
 and Peter V. Coveney
\footnote{\small{\texttt{p.v.coveney@ucl.ac.uk}}\normalsize}	\\
}
\affiliation{\small{Centre for Computational Science, Department of
Chemistry, University College London, 20 Gordon Street, London WC1H
0AJ, UK}}

\vspace{1.5cm}
\date{\today} 
\vspace{2.5cm}
\begin{abstract} 
By means of a three-dimensional amphiphilic lattice-Boltzmann model with 
short-range interactions for the description of ternary amphiphilic
fluids, we study how the phase separation kinetics of a symmetric
binary immiscible fluid is altered by the presence of the amphiphilic
species. We find that a gradual increase in amphiphile concentration
slows down domain growth, initially from algebraic, to logarithmic
temporal dependence, and, at higher concentrations, from logarithmic
to stretched-exponential form. In growth-arrested
stretched-exponential regimes, at late times we observe the self-assembly of 
sponge mesophases and gyroid liquid crystalline cubic mesophases,
hence confirming that (a) amphiphile-amphiphile interactions need not
be long-ranged in order for periodically modulated structures to arise
in a dynamics of competing interactions, and (b) a chemically-specific
model of the amphiphile is not required for the self-assembly of cubic 
mesophases, contradicting claims in the literature. We also observe a
structural order-disorder transition between sponge and gyroid phases
driven by amphiphile concentration alone or, independently, by the
amphiphile-amphiphile and the amphiphile-binary fluid coupling
parameters. For the growth-arrested mesophases, we also observe
temporal oscillations in the structure function at all length scales;
most of the wavenumbers show slow decay, and long-term stationarity or
growth for the others. We ascribe this behaviour to a combination of
complex amphiphile dynamics leading to Marangoni flows. 
\end{abstract}
\pacs{mm.nn.xx}


\maketitle

\newpage

\section{Introduction}

The term {\em amphiphilic fluid} is broadly used to denote multiphase
fluids in which at least one species is of a surfactant nature. A 
surfactant molecule (from {\em surf}ace {\em act}ive age{\em nt},
which we shall also refer to as an amphiphile)
contains a polar headgroup attached to a hydrocarbon tail which,
dispersed in a binary immiscible fluid mixture, such as oil and water,
is driven towards and adsorbed at the interface between the two
fluids. The selective chemical affinity between each part of the
surfactant molecule and the components of the binary mixture is
the mechanism responsible for such a taxis \cite{GELBART}. Not only are
amphiphilic fluids important in physical chemistry, structural biology,
soft matter physics and materials science from a fundamental perspective,
but their applications are also widespread. Detergents and 
mammalian respiration are two common examples in which surfactants are 
present. Living cell membranes are complex macromolecular assemblies 
comprised in large part of self-assembled phospholipids, of an 
amphiphilic nature \cite{AMPHI_BIO}. Sponge mesophases are formed as a 
result of an amphiphile dispersion or melt at an appropriate composition, 
and enjoy numerous applications in medical research as well as the 
pharmaceutical, cosmetic, food, and agro- and petrochemical industries 
\cite{MICROEMUL_INDUST,GOMPPER&SCHICK2}. Liquid-crystalline
bicontinuous cubic mesophases of monoglycerides and the lipid extract
from archaebacterium {\em Sulfolobus solfataricus} have been found at
physiological conditions in cell organelles and physiological
transient processes such as membrane budding, cell permeation and the
digestion of fats \cite{MARIANI}. Amphiphilic cubic mesophases can
also be synthesised   
for important applications in membrane protein crystallisation,
controlled drug release and biosensors \cite{MARRINK,LUZZATI}. These
phases are termed {\em mesophases} not only because their intrinsic
internal length scales range between those characteristic of molecular
and hydrodynamic (or macroscopic) realms, but also their mechanical 
properties are half-way between those found in a liquid and a solid
\cite{GELBART,AMPHI_BIO,GOMPPER&SCHICK3}.  

Amphiphiles have the property of lowering the interfacial tension in a
{\em }binary immiscible, say oil-water, fluid~\cite{GOMPPER&SCHICK3}. 
Given the bipolar nature of 
their molecular structure, amphiphile adsorption at the oil-water
interface is a process which is energetically favoured relative to their
entropically beneficial dispersion in the bulk. This effectively 
reduces the pressure tensor at the interface, making the immiscible 
species more alike. As more interfacial surface is created, so more 
amphiphile dispersed in the bulk can be accommodated at it. 

The effect of adding surfactant above a critical concentration to an 
oil-water mixture undergoing phase separation is to slow down the 
demixing process, which, with the addition of sufficient amphiphile, 
can be totally arrested. Langevin, molecular dynamics and lattice gas
simulations have shown that, as the concentration of dispersed surfactant 
increases, the temporal growth law of the average size of the 
immiscible oil-water domains, of the power-law form $t^a$ in the
surfactantless case \cite{P2,BRAY}, is seen to cross over to a slower,
logarithmic growth of the form $(\ln t)^\theta$, where $a$ and
$\theta$ are fitting parameters and $t$ is the time
\cite{KAWAKATSU93,LARADJI92,EMERTON97}. Emerton~{\em et~al.} 
showed that increasing the surfactant concentration even 
further leads to growth well described by the stretched exponential
$A-B\exp(-Ct^D)$, where $A$, $B$, $C$ and $D$ are fitting parameters,
including halted segregation at sufficiently late times
\cite{EMERTON97}. Depending on temperature, pressure and fluid 
composition in such a stretched exponential regime, the amphiphile can 
self-assemble and force the oil-water mixture into a wealth of 
equilibrium structures. The self-assembling process is dictated by the 
competing attraction-repulsion mechanisms present among the species. 
Lamellae and hexagonally-packed cylinders are examples of
these mesophases, also referred to as $L_\alpha$ and $H$, respectively,
with continuous translational symmetry along one or two
directions. Other examples are the sponge ($L_3$) mesophase and the
micellar  
(Q$^{223}$ or $Pm3n$, and Q$^{227}$ or $Fd3m$), primitive 
(``P", Q$^{229}$ or $Im3m$), diamond (``D", ``F", Q$^{224}$ or $Pn3m$) 
and gyroid (``G", Q$^{230}$ or $Ia\overline{3}d$) cubic mesophases,
all of which lack continuous translational symmetry
\cite{SEDDON}. Among all the aforementioned phases, only the sponge
mesophase is devoid of long-range order and so cannot be classified as
a liquid crystal: it is rather characterised by glassy features. 

A sponge mesophase formed by the amphiphilic stabilisation of a
phase-segregating binary fluid mixture is called a
microemulsion. Since we shall be  
dealing with oil and water in equal proportions, we shall be concerned
with bicontinuous microemulsions. A bicontinuous microemulsion is 
a structure consisting of two percolating, interpenetrating oil and 
water phases separated by a monolayer of surfactant molecules 
adsorbed at the interface. Oil and water are isotropically mixed, and
ordering is short range. Sponge phases formed by the dispersion of
amphiphile in a single phase solvent differ from microemulsions in
that it is a surfactant bilayer which underlies the structure, and the
regions it divides are occupied by the same fluid component. A gyroid 
phase is also a bicontinuous, interpenetrating structure; however, 
ordering is evidently long range, whence its classification as a liquid
crystal. In the gyroid, the locus where most of the surfactant resides is 
a triply periodic minimal surface (TPMS) whose unit cell is of cubic 
symmetry. The surface has zero mean curvature, no two points on it are 
connected by a straight segment, and no reflexion symmetries are present. 
Isosurfaces of the gyroid phase for which oil and water are not at equal 
composition (minority phases) form mutually percolating, three-fold 
coordinated, regular lattices. Other examples of triply periodic surfaces 
of zero mean curvature arise in the P and D mesophases, the minority 
phase isosurfaces of which exhibit coordination numbers of six and four, 
respectively.   

The purpose of the present paper is to report on a theoretical study of the 
segregation kinetics in ternary amphiphilic fluids and the 
self-assembly of the sponge and gyroid mesophases. By progressively 
adding surfactant to an initially homogeneous immiscible
oil-water mixture on the way to achieving arrested domain
growth, we shall give an account of how the segregation kinetics of
the fluid domains is affected by the addition of surfactant, and study
the features of the associated mesophases  that are formed. The
mesophases corresponding to such late time, arrested  growth regimes
are sponges which turn into gyroids as we increase the  surfactant
concentration. We shall also see that these phases exhibit temporal
oscillations in the size of the oil-water domains, which we ascribe to
Marangoni flows.

\section{Overview of modelling and simulation of amphiphilic fluid
self-assembly} 

Various methods have been used to date to model and simulate ternary
amphiphilic mixtures and to study their phase segregation kinetics and
the formation of microemulsions and liquid crystalline phases. We
briefly review them in this section.

Kawakatsu~{\em et~al.} studied segregation kinetics employing a two 
dimensional hybrid model with thermal noise but without hydrodynamics, 
combining a continuum, Langevin diffusion equation for the oil-water
dynamics and Newtonian dynamics with dissipation for bipolar particles 
modelling the surfactant \cite{KAWAKATSU93}. They used a
free energy in the form of a $\phi^4$-Ginzburg-Landau expansion
\cite{HOHENBERG} plus terms modelling the surfactant-interface and
surfactant-surfactant interactions. They found the average domain size 
of symmetric binary immiscible fluids with amphiphile to grow with
time more slowly than $t^{1/3}$, the latter expected for binary alloys
in two and three dimensions. Laradji~{\em et~al.}, instead of
modelling the amphiphile as a particulate species, regarded it as a
continuous density coupled to the
oil-water order parameter in a $\phi^4$-Ginzburg-Landau free energy
\cite{LARADJI92}. In their work, they studied several cases of 
two-dimensional Langevin diffusion equations, one of which being the
so-called Model~D \cite{HALPERIN}.
Model~D incorporates noise, a conserved order parameter and surfactant 
density, but excludes hydrodynamics. Laradji~{\em et~al.} not only found 
logarithmic growth for the behaviour of the average domain size with time, 
but also observed a slowdown from it for higher surfactant concentrations
and dynamical scaling for the structure function at intermediate times. 
Yao~\&~Laradji, using a modified Lifshitz-Slyozov nucleation
theory for continuum fields in two and three dimensions, studied how the 
Ostwald ripening dynamics of an asymmetric mixture of oil and water is 
altered by the presence of a surfactant species \cite{YAO}. They found 
results similar to those of Laradji~{\em et~al.} \cite{LARADJI92}.  

The segregation kinetics of amphiphilic fluids have also been studied
with fully particulate methods such as classical molecular dynamics
and, more recently, hydrodynamic lattice gases. Using a minimalist
molecular dynamics model in two dimensions, Laradji~{\em
et~al.}~\cite{LARADJI94} found a crossover scaling function similar to
previous Langevin \cite{KAWAKATSU93} and Lifshitz-Slyozov models
\cite{YAO}, yet with a different algebraic exponent, and a slowing
down from the algebraic growth laws for binary mixtures. Using
two-dimensional hydrodynamic lattice gas models for symmetric
\cite{EMERTON97} and asymmetric mixtures \cite{WEIG97}, the group of
Coveney found that surfactant induces a crossover to a logarithmic
slow growth, and, with sufficient surfactant, full arrest of domain
growth which is well described by a stretched exponential
function. The group found similar results with a three-dimensional
hydrodynamic lattice gas model~\cite{LOVE01}. 

Particulate methods have also been used to tackle
mesophase self-assembly. Using classical molecular dynamics methods,
Marrink~{\em et~al.} simulated evolution of a surfactant bilayer, 
initially set up on the morphology of a ``D" TPMS, to study both the
surfactant packing structure and how close such a bilayer would remain
to the TPMS after relaxation ~\cite{MARRINK}. They, however, did not
address self-assembly dynamics: time scales required for that are
orders of magnitude above those reachable with atomistic techniques on
present-day cutting-edge supercomputers. In dissipative particle dynamics 
(DPD) approaches, a Langevin dynamics with momentum conservation is
solved to model ill-defined, mesoscopic dissipative particles
interacting via repulsive, soft potentials; hydrodynamics is emergent
and the amphiphile is represented by dissipative particles bound 
together by rods or springs \cite{GROOT2,GROOT,PRINSEN}. The DPD
simulations of Groot \& Madden of copolymer melts \cite{GROOT2} showed
that melts of symmetric amphiphile led to lamellar phases, whereas a
gyroid-like structure appeared only for asymmetric amphiphile as a
transient phase, precursor to a hexagonally packed tubular
phase. Nekovee \& Coveney, using the lattice-Boltzmann model we employ
in this work, were able to reproduce the ``P" mesophase in a binary
amphiphilic mixture of surfactant and solvent \cite{JACS_P}. 

Many of the simulation studies on the formation kinetics of
microemulsion and liquid crystalline mesophases have made use of
stochastic Langevin diffusion methods, in which mass currents are
driven by chemical potential gradients computed from free 
energies of the ubiquitous $\phi^4$-Ginzburg-Landau expansion
form. These models treat the amphiphile only implicitly through the
functional dependence of the surface tension parameter with the
amphiphile density~\cite{GOMPPER2,GOMPPER,NONOMURA,IMAI,QI}. In cases
in which the amphiphile is a copolymer, however, the free energy is
often derived from polymer models which aim at accounting for the
amphiphile's molecular structure with a certain degree of
specificity~\cite{OHTA,ZVELINDOVSKY,VLIMMEREN}. The validity of these
Flory-Huggins type approaches rests on being able to derive the free
energy from a microscopic model of the complex fluid mixture, which
not only might entail considerable difficulty but does require the
segregation to be a quasi-static, local equilibrium process. Under
general far from equilibrium conditions, such as occurs in the
sudden-quench scenario so often employed in the literature,
equilibrium thermodynamic potentials are known not to adequately
describe the process. Besides, free energy based methods also require
surfactant adsorption and relaxation on the interface to be much
faster than interface motion, a so-called {\em adiabatic
approximation}. Free energy approaches are frequently  
represented as paradigms of thermodynamically consistent mesoscopic
methods; some of them also pursue chemical specificity in elaborate
empirical exercises amounting to little more than parameter
fitting of polymer models. The philosophy behind them, nonetheless, is
the use of macroscopic, local equilibrium information to specify a
stochastic, and hence mesoscopic, non-equilibrium dynamics. None of 
these methods offers a dynamics satisfying detailed balance, let alone
an $H$-theorem (Lyapunov function) guaranteeing irreversible evolution
towards the equilibrium state described by the prescribed free
energy. As a consequence, the `thermodynamic consistency' of these
methods remains on shaky grounds.  

The fact that some free energy approaches \cite{ZVELINDOVSKY,VLIMMEREN} 
focus on the specific molecular structure of the amphiphile raises the 
question of what use particulate methods, such as is the one we report
in this paper, have in the 
simulation of amphiphilic fluid systems. Our method, by reducing
the description of the amphiphilic molecule to its minimal possible
expression---a point dipole, retains the minimum number of degrees of
freedom necessary to model interfacial adsorption and
micellisation, and, additionally, in a hydrodynamically consistent
framework which does not require processes to be quasi-static. With
these basic properties 
at our disposal, we want to fully exploit our model's capabilities to
determine the non-equilibrium amphiphilic dynamics and the equilibrium 
fluid structures arising from it. The minimalistic bottom-up
approach is in line with the fact that, far enough from criticality, 
distinct molecular structures and microscopic dynamics can produce
similar macroscopic behaviour---this is universality emerging
from microscopic complexity \cite{COVNOBEL}. In addition, particulate
methods are much more suitable than conventional continuum fluid
dynamics methods \cite{NADIM} for the simulation of interface
dynamics. Such dynamics is an emergent property of the underlying
inter-particle interactions among the immiscible species; a set of
continuum partial differential equations describing the locus of the
interface is, rather, its macroscopic manifestation, and its solution
a much more labourious endeavour. {\em A fortiori}, modelling
surfactant adsorption and self-assembly in an explicit fashion via
particulate methods provides a more realistic picture of the
microscopics than doing so at the continuum, macroscopic
limit described by free-energy approaches.

Lattice-Boltzmann (LB) methods were originally developed as a means of
reducing the computational cost associated with lattice-gas automaton
(LGA) algorithms \cite{SUCCI}. LB methods evolve a single-particle 
distribution function via a discretised Boltzmann equation, usually in
the linearised, relaxation-time (BGK) approximation. Such a
single-particle distribution, at a particular 
time slice and spatial position on the lattice, is an average over the
LGA velocity space for a statistically large number of different
microscopic realisations (initial conditions). The fact that much of
the phenomenology of binary immiscible and ternary amphiphilic fluids 
occurs for small spatio-temporal gradients permits us to take the
mean-field (or molecular chaos) approximation, and the Boltzmann-Grad 
limit in which such an approximation holds, as heuristically
appropriate for the modelling of their universal properties. 
Heuristics come into play in that tunable parameters are introduced in 
LB models in order to reproduce desired quantities of dense and/or
complex fluids, such as surface tension, viscosity and thermal
conductivity (for required values of Reynolds and Prandtl numbers),
stress tensors (for required viscous or viscoelastic behaviour), and
equations of state (for liquid-gas and phase segregating
transitions). It is worth noting that the increasing popularity of LB
methods in recent years is primarily based on pragmatic
considerations associated with their simplicity and algorithmic
efficiency.  

This paper presents the first quantitative account of amphiphilic
phase segregation dynamics using a three-dimensional model based on
the Boltzmann transport equation. It describes the spontaneous
self-assembly of the gyroid liquid crystalline cubic mesophase and an
order-disorder transition between the latter and the sponge mesophase,
of glassy features. The remainder of the paper is structured as
follows. In Section III we describe the model. In Section IV we look
at how the  segregation kinetics of the fluid domains is affected by
the addition of surfactant. Section V studies the temporal
oscillations of the 
average domain size and the structure function, which are only
observed for segregation-halted regimes. In Section VI we characterise
the morphology of the mesophases corresponding to those regimes via
direct- and Fourier-space imaging, and identify the
sponge$\leftrightarrow$gyroid structural transition. Finally, we 
provide conclusions in Section VII.

\section{A lattice-Boltzmann model for ternary amphiphilic fluids}
\label{SCLB}

The amphiphilic lattice-Boltzmann model we employ in this paper is
derived from that originally proposed by Chen~{\em
  et~al.}~\cite{CHEN_AMPHI,MAZIAR-ET-AL}. The method can be regarded
as a fully mesoscopic, bottom-up approach, which does not require the
existence of a thermodynamic potential describing phase
transitions. In fact, the method is athermal in the sense that, for
algorithmic efficiency reasons, the microdynamics is devised {\em
ex~professo} to conserve velocity moments of the distribution function
only up to first-order; this simplification is valid wherever thermal
fluctuations are negligible, e.g. away from criticality. This is, for
example, the case of deep quenches into the spinodal region of the
fluid's phase diagram, which is our case in this paper. 
As opposed to top-down LB methods, based
on the imposition of a free energy functional~\cite{LAMURA,THEISSEN},
the global dynamics arise as 
an emergent property of the interactions between mesoscopic levels of
description, in agreement with a complexity paradigm
\cite{COVNOBEL}. Oil-water segregation is achieved via inter-species
forces which modify the fluid's macroscopic velocity. The dynamics in
the bulk of each binary immiscible species (e.g. oil and water) can be
derived from a Boltzmann equation with a forcing term. An amphiphilic
molecule is modelled as a continuously orientable massive point dipole
subjected to thermal noise and relaxing towards an equilibrium that
minimises its interaction energy with mean fields generated by its
nearest neighbours on the lattice. The densities of surfactant, oil
and water evolve via coupled lattice-BGK equations. This is a
mean-field approach which exhibits Galilean invariance and reproduces
correct hydrodynamics. We have also shown, in a previous paper which
serves as a reference benchmark for this study 
\cite{P2}, that the model reproduces the dynamical scaling hypothesis
during the phase segregation experienced by binary immiscible
(oil-water) fluids. Its algorithmic simplicity allows it to
achieve extremely high performance on massively parallel computers
\cite{COMPARISON}, and substantially reduces the domain of numerical
instability present in free-energy-based LB methods \cite{P2}. Because
an $H$-theorem is lacking in essentially all multiphase
lattice-Boltzmann models hitherto proposed \cite{ENTROPIC}, we
consider it artificial to try to enforce a prescribed thermodynamic
equilibrium in these schemes; a method which is algorithmically
simpler, fully mesoscopic, mean-field and bottom-up is of greater
fundamental interest.

\subsection{Binary immiscible fluids}
\label{BINARY-IMMISC-LBE}
The core of our model is a lattice-BGK equation governing the
evolution of the mass density distribution $m^\alpha
n_k^\alpha(\mathbf{x},t)$ of component $\alpha$ in an interacting
fluid mixture at position $\mathbf{x}$, instant $t$, and for discrete
molecular velocity $\mathbf{c}_k$, on a regular lattice and in discrete
time. Here, $m^\alpha$ is the particle mass which we set to unity for
convenience, and the single-particle distribution
$n_k^\alpha(\mathbf{x},t)$ obeys the lattice-BGK relaxation-streaming
mechanism:   
\begin{equation}
	n_k^\alpha(\mathbf{x + c}_k,t+1) - n_k^\alpha({\mathbf x},t)
	= \Omega_k^\alpha \,,
\label{LBEQUATION}
\end{equation}
\noindent where the collision term has two contributions accounting
for the kinetics of non-interacting (ideal) plus interacting 
(non-ideal) multicomponent species, respectively:
\be
	\Omega_k^\alpha({\mathbf x},t)
	\equiv
	\Omega_k^{(0)\alpha}({\mathbf x},t)
	+
	\sum_{\bar\alpha}\sum_l  \Lambda_{kl}^{\alpha\bar\alpha}
	n_l^{\bar\alpha}				\,.
	\label{COLLISION-TERM}
\ee
\noi the sums extending over all available species and directions,
and 
\be
	\Omega_k^{(0)\alpha}({\mathbf x},t) 
	\equiv
	-\frac{n_k^\alpha({\mathbf x},t) 
	- n_k^{\alpha(\mathrm{eq})}({\mathbf x},t)}
	{\tau^\alpha}						
	\label{IDEAL-COLLISION-TERM}
	\,.
\ee 
\noindent Here, the time increment and lattice spacing are both unity,
${\mathbf x}$ is a node of such a lattice, 
$\alpha=\mathrm{r},\mathrm{b}$ (e.g. oil (r) or water (b)), and
${\mathbf c}_k$ is one of the 24 ($\equiv N_\mathrm{vec}$) discrete
velocity vectors plus one null velocity of the projected face centred
hypercubic D4Q25 lattice we use to guarantee isotropy in the 
macroscopic equations that the model reproduces for a bulk, single
phase fluid \cite{FRISCH}. The parameter $\tau^\alpha$ defines a
single relaxation rate towards equilibrium for component $\alpha$,
$\Lambda_{kl}^{\alpha\bar\alpha}$ can be regarded as a matrix element
of a cross-collision operator $\Lambda$ which is a function of both
$\tau^\alpha$ and the acceleration $\mathbf{a}^\alpha$, the latter
being experienced by a fluid element due to its neighbours, as 
will be defined later. The function 
$n_k^{\alpha(\mathrm{eq})}({\mathbf x},t)$ in 
Eq.~(\ref{IDEAL-COLLISION-TERM}) is the discretisation of a
third-order expansion in Mach number of a local Maxwellian \cite{P2},
representing the local equilibrium state of the $\alpha$th component, 
\begin{equation}
	n_k^{\alpha(\mathrm{eq})}({\mathbf x},t)
	= 
	\omega_k \, n^\alpha({\mathbf x},t)
	\Big[ 1 +
	\frac{1}{c_s^2}{\mathbf c}_k\cdot{{\mathbf u}} +
	\frac{1}{2c_s^4}({\mathbf c}_k\cdot{{\mathbf u}})^2 -
	\frac{1}{2c_s^2}u^2 + \frac{1}{6c_s^6}({\mathbf
	c}_k\cdot{{\mathbf u}})^3 - \frac{1}{2c_s^4}u^2({\mathbf
	c}_k\cdot{{\mathbf u}}) \Big]\,,
\label{EQUIL}
\end{equation}
\noindent where $\omega_k$ are the coefficients resulting from the
velocity space discretisation, and  $c_s$ is the speed of sound, both
of which are determined by the choice of the lattice. For the
projected D4Q25 lattice we use, the speed of sound is
$c_s=1/\sqrt{3}$, $\omega_k=1/3$ for the speed $c_k=0$, and $1/36$ for
speeds $c_k=1$ and $\sqrt{2}$. In Eq.~(\ref{EQUIL}),
$\mathbf{u}=\mathbf{u}(\mathbf{x},t)$ is the macroscopic
velocity of the mixture, through which the collision term couples the
different molecular velocities ${\mathbf c}_k$. This is because
$\mathbf{u}$ is a function of the components' macroscopic velocities,
defined as $n^\alpha({\mathbf x},t)\,{\mathbf u}^\alpha\equiv \sum_k
n_k^\alpha({\mathbf x},t)\,{\mathbf c}_k$. 

A judicious choice of the coefficients in the expansion of the
equilibrium distribution Eq.~(\ref{EQUIL}) allows for mass and momentum to 
be (non-locally) conserved for the non-interacting, ideal gas mixture case, 
i.e.
\begin{equation}
	\sum_k\Omega_k^{(0)\alpha}=0			\,,\qquad	
	\sum_\alpha 
	m_\alpha 
	\sum_k{\mathbf c}_k\Omega_k^{(0)\alpha}=0	\,. 	 
	\label{BGK-CONSERVATIONS}
\end{equation}
\noindent It can be shown that in the limit of creeping flows to
second order, i.e. $u^2 \approx 0$, the expression for the fluid
mixture's macroscopic velocity ${\mathbf u}$ required for momentum
conservation in the absence of interactions, as a function of
${\mathbf u}^\alpha$, simplifies to that obtained for a second-order
expansion of the equilibrium distribution, namely ${\mathbf u}
\equiv\sum_\alpha\frac{\rho^\alpha{\mathbf u}^\alpha}{\tau^\alpha}/
\sum_\alpha\frac{\rho^\alpha}{\tau^\alpha}$, which we have
incorporated in our implementation.

The form of the collision term (\ref{COLLISION-TERM}) derives from
adding an increment $\Delta{\mathbf u}^\alpha$ to the fluid mixture's
macroscopic velocity ${\mathbf u}$ which enters in the equilibrium
distribution (\ref{EQUIL}), i.e. 
$\Omega_k^\alpha(\mathbf{u})\equiv\Omega_k^{\mathrm{(0)}\alpha}
(\mathbf{u}+\Delta{\mathbf u}^\alpha)$ 
where 
$\Delta{\mathbf	u}^\alpha\equiv\mathbf{a}^\alpha\tau^\alpha$ and
${\mathbf a}^\alpha\equiv{\mathbf F}^\alpha/\rho ^\alpha$. Here
\begin{equation}
	{\mathbf F}^{\alpha ,\mathrm{c}}({\mathbf x},t)
	\equiv
	-\psi^\alpha({\mathbf
	x},t)\sum_{\bar\alpha} g_{\alpha\bar\alpha} \sum_{{\mathbf
	x}'}  \psi^{\bar\alpha}({\mathbf x}',t) ({\mathbf x}'-{\mathbf
	x}) \label{LBE-FORCE}
\end{equation}
\noi is the mean-field force density felt by phase $\alpha$ at site
${\mathbf x}$ and time $t$ due to its surroundings;
$g_{\alpha\bar\alpha}$ is a coupling matrix controlling 
the interfacial tension between the fluid species, interface
adsorption/desorption properties of the surfactant molecules, and the
surfactant-surfactant interaction; $\psi^\alpha$ is an {\em effective
mass} which serves as a functional parameter and can have a variety of
forms for modelling various types of fluids. We only allow
nearest-neighbour interactions,  
${\mathbf x}'\equiv{\mathbf x}+{\mathbf c}_k$, and choose
$\psi^\alpha({\mathbf x},t)\equiv 1-\exp\Big[-n^\alpha({\mathbf
x},t)\Big]$, where $n^\alpha\equiv\sum_k n_k^\alpha$. This choice for
$\psi$ has also been made by Shan and Chen to model liquid-gas phase
transitions \cite{SHAN-CHEN} although, as we shall see, our
motivation here is different.

\subsection{Amphiphilic fluids}

The incorporation of a third, amphiphilic species not only requires
the inclusion of an extra variable (``s'') for the superscript
denoting the species in Eq. (\ref{LBEQUATION}), but also a
modification of the cross-collision operator $\Lambda$ since
amphiphiles interact with fluid elements and between themselves. In
addition, the physics of amphiphilic molecules, namely, self-assembly 
and adsorption to immiscible fluid interfaces, cannot be modelled
without introducing a new type of body force: in Subsection  
\ref{BINARY-IMMISC-LBE} ordinary bulk fluid species are thought of as
point-like particles given that their interactions depend on their
relative distances alone. For surfactant molecules, however, their
orientations are important too \cite{CHEN_AMPHI}, and a dipole is the 
simplest configuration to mimic their essential character. In short,
we must extend the scalar lattice-BGK model hitherto described
into a vector model. 

Each surfactant molecule is represented by an average dipole vector 
${\mathbf d}({\mathbf x},t)$ at each site and time step, whose
orientation is allowed to vary continuously.  The average is taken
over nearest neighbours before advection, according to the propagation 
equation 
\be
	n^{\mathrm s}({\mathbf x},t+1){\mathbf d}({\mathbf x},t+1)
	=
	\sum_k
	\tilde{n}^{\mathrm s}_k(\mathbf{x - c}_k,t)
	\tilde{{\mathbf d}}(\mathbf{x - c}_k,t)			\,,
	\label{SURF-PROPAGATION}
\ee
\noindent where the tildes denote post-collisional values, as defined
by Eq. (\ref{LBEQUATION}) for the $\Lambda\equiv 0$
($g_{\alpha\bar\alpha}\equiv 0$) case if we replace the leftmost
summand with $\tilde{n}^\alpha_k({\mathbf x},t)$. For the sake of
simplicity and computational efficiency, the model does not assign
microscopic velocities ${\mathbf c}_k$ to single dipole vectors but to
site-averaged surfactant densities instead, as can be seen, for
example on the right hand side of Eq.~(\ref{SURF-PROPAGATION}). 

Dipole relaxation is governed by the BGK process
\be
	\tilde{{\mathbf d}}({\mathbf x},t)
	=
	{\mathbf d}({\mathbf x},t)
	-
	\frac{1}{\tau^\mathrm{s}}\Big[
		{\mathbf d}({\mathbf x},t) 
		- {\mathbf d}^{\mathrm{eq}}({\mathbf x},t)
	\Big]							\,,
\ee
\noindent where $\tau^\mathrm{s}$ is a new parameter controlling
the relaxation towards the local equilibrium ${\mathbf
d}^{\mathrm{eq}}({\mathbf x},t)$, which is understood as the average
orientation with respect to the Gibbs measure, i.e. 
\be
	{\mathbf d}^{\mathrm{eq}}({\mathbf x},t)
	\equiv d_0 \,
	\frac{\int \mathrm{d}^2\Omega\,\,
		{\mathrm e}^{-\beta H_{\hat\Omega}({\mathbf x},t)}
		\hat{\Omega}}
		{\int \mathrm{d}^2\Omega\,\,
		{\mathrm e}^{-\beta H_{\hat\Omega}({\mathbf x},t)}}
	\label{DIPOLE_EQUIL}\,,
\ee
\noi where $\mathrm{d}^2\Omega$ is an element of solid angle whose
director is the unit vector $\hat\Omega$ representing the dipole
orientation, and $\beta$ is an inverse temperature-like
parameter. The modulus of the distribution (\ref{DIPOLE_EQUIL}) ranges
between 0 and the scale value $d_0$, chosen to be unity for
convenience. That, along with $\tau^\mathrm{s}\ge 1$, guarantee the
magnitude of the dipole vector to be less than $d_0$ at all
times. Equation (\ref{DIPOLE_EQUIL}) favours surfactant orientations which
minimise the {\em energy}
$H_{\hat\Omega} \equiv -\hat\Omega \cdot\mathbf{h}({\mathbf x},t)$, 
where $\mathbf{h}({\mathbf x},t)$ is the sum of the mean fields
created by surrounding bulk fluid and surfactant, namely 
\bea
	\mathbf{h}^\mathrm{c}({\mathbf x},t)
	&\equiv&
	\sum_\alpha q_\alpha 
	\sum_k n^\alpha(\mathbf{x+c}_k,t){\mathbf c}_k
								\,,\\
	\mathbf{h}^\mathrm{s}({\mathbf x},t)
	&\equiv&
	\sum_k\Big[
		n_k^\mathrm{s}({\mathbf x},t)
		\mathbf{d}({\mathbf x},t)	+
		\sum_{l\ne 0}n^\mathrm{s}(\mathbf{x+c}_l,t)
		\mathbf{\theta}_l\cdot
		\mathbf{d}(\mathbf{x+c}_l,t)\Big]		\,.
\eea
\noi allowing for nearest-neighbour interactions only. The first
equation is a discrete approximation to the colour 
gradient for the immiscible species, where $q_\alpha=0,\pm 1$ is the
colour charge of species $\alpha$. The second equation is a dipole
vector density, where summation over $k$ performs local dipole
averaging, summation over $l$ includes all nearest neighbour
contributions, and the second-rank tensor 
$\mathbf{\theta}_l \equiv \mathbf{I}-D\mathbf{c}_l\mathbf{c}_l/c^2$,
where $c$ is the modulus of $\mathbf{c}_l$ and $D$ is the spatial
dimension, picks up desired orientations from
nearest-neighbour dipoles. Finally, Eq. (\ref{DIPOLE_EQUIL}) can be
integrated analytically in three dimensions to give
$\mathbf{d}^\mathrm{eq}=d_0 \Big[\coth(\beta h)-\frac{1}{\beta
h}\Big]\hat{\mathbf{h}}$, where $h$ is the magnitude of $\mathbf{h}$
and $\hat{\mathbf{h}}$ its unit vector.  

The new interactions that modify the interspecies collision operator
$\Lambda$ are the force on an immiscible fluid element from other
fluid elements and amphiphiles, $\mathbf{F}^\alpha \equiv
g_\mathrm{br}\mathbf{F}^\mathrm{\alpha,c} +
g_\mathrm{bs}\mathbf{F}^\mathrm{\alpha,s}$, where
$\mathbf{F}^\mathrm{\alpha,c}$ is that in Eq. (\ref{LBE-FORCE}), and
the force on an amphiphilic molecule from neighbouring fluid elements
and amphiphiles, $\mathbf{F}^\mathrm{s} \equiv
g_\mathrm{bs}\mathbf{F}^\mathrm{s,c} +
g_\mathrm{ss}\mathbf{F}^\mathrm{s,s}$. In these expressions,
$g_\mathrm{br}$, $g_\mathrm{bs}$ and $g_\mathrm{ss}$ are coupling
scalar parameters, and the analytical expressions for each force term,
derived in \cite{CHEN_AMPHI}, are
\bea
	\mathbf{F}^\mathrm{\alpha,s}
	&\equiv&
	-2g_\mathrm{\alpha s}\psi^\alpha({\mathbf x},t)
	\sum_{k\ne 0}
	\tilde{\mathbf{d}}(\mathbf{x+c}_l,t)
		\mathbf{\theta}_l\cdot
		\psi^\mathrm{s}(\mathbf{x+c}_l,t)		
						\label{SURF_FORCE1}\,\\
	\mathbf{F}^\mathrm{s,c}
	&\equiv&
	2\psi^\mathrm{s}({\mathbf x},t)
	\tilde{\mathbf{d}}(\mathbf{x+c}_l,t) \cdot
	\sum_\alpha g_\mathrm{\alpha s}
	\sum_{k\ne 0}\mathbf{\theta}_k
	\psi^\alpha(\mathbf{x+c}_k,t)
						\label{SURF_FORCE2}\,\\
	\mathbf{F}^\mathrm{s,s}
	&\equiv&
	-\frac{4D}{c^2}g_\mathrm{ss}\psi^\mathrm{s}({\mathbf x},t)
	\sum_{k}
	\Big\{
	\tilde{\mathbf{d}}(\mathbf{x+c}_k,t) \cdot \theta_k \cdot 
	\tilde{\mathbf{d}}(\mathbf{x},t) \mathbf{c}_k	
							\nonumber\\
	&+&							
	\Big[\tilde{\mathbf{d}}(\mathbf{x+c}_k,t)
	\tilde{\mathbf{d}}({\mathbf x},t) +
	\tilde{\mathbf{d}}({\mathbf x},t)
	\tilde{\mathbf{d}}(\mathbf{x+c}_k,t)\Big]
	\cdot\mathbf{c}_k
	\Big\}
	\psi^\mathrm{s}(\mathbf{x+c}_k,t)	
						\label{SURF_FORCE3}\,.
\eea
\noi Equations (\ref{SURF_FORCE1}), (\ref{SURF_FORCE2}), and
(\ref{SURF_FORCE3}) were derived considering only nearest-neighbour
interactions, modelling each dipole as a dumbell of oppositely
colour-charged particles displaced $\pm\mathbf{d}/2$ from the dipole's
centre of mass location $\mathbf{x}$, and carrying out Taylor
expansions of the force (\ref{LBE-FORCE}) to leading order in
$\mathbf{d}$ about $\mathbf{x}$ as well as those at the neighbouring
sites \cite{CHEN_AMPHI}. Also, Eq. (\ref{SURF_FORCE2}) is the reaction to
force (\ref{SURF_FORCE1}), and Taylor expansions in the ratio of
$|\mathbf{c}_k|$ to the length scale that the colour gradient sets can
be used to further simplify the expressions. Finally, additional
coupling parameters $g_{\alpha\bar\alpha}$ have been introduced, where
$g_\mathrm{ss}$ should be chosen negative to model attraction between
two amphiphile heads or tails, and repulsion between a head and a
tail.

\subsection{Selection of the parameters for the simulations}
\label{SIMUL}

The model is implemented as a parallel code in Fortran90 making
use of the Message Passing Interface parallel paradigm \cite{MPI} and
spatial domain decomposition, and incorporating wrap-around, periodical 
boundary conditions. It was executed on 16 to 64 processors on 
SGI Origin2000 and Origin3800 parallel platforms. The form 
$\psi\equiv 1-\exp[-n(\mathbf{x},t)]$ for the effective mass in the
force in Eq.~(\ref{LBE-FORCE}) was heuristically chosen so as to broaden
the region of numerical stability in parameter space: numerical
instabilities can arise as the result of high values of forces and
speeds, and are more likely to occur in our model when surfactant
interactions are included than for binary immiscible fluids
\cite{IAIN, P2}.

Preliminary studies allowed us to determine the values of the model's
various parameters for which an initially thorough mixture of two 
immiscible fluid phases plus a dispersed amphiphilic species produced
a segregated mixture with arrested domain growth \cite{IAIN}. Those 
values were the surfactant thermal parameter $\beta=10.0$, all particle
masses and relaxation times set to 1.0, and coupling constants 
$g_{\mathrm{br}}=0.08$, $g_{\mathrm{bs}}=-0.006$, and
$g_{\mathrm{ss}}=-0.003$. (Masses, $m^\alpha$, enter in the description
through $\rho^\alpha(\mathbf{x},t)\equiv \sum_k m^\alpha
n_k^\alpha(\mathbf{x},t)$.) We simulated the behaviour of a ternary 
mixture by varying the coupling constants around the values mentioned 
above, and for initial surfactant particle densities ranging in the 
interval $0.00\le n^\mathrm{(0)s}\le0.90$. The lattice sites and
directions were initially populated with flatly distributed mass
densities, $0 \le \rho_k^\alpha(\mathbf{x},0) \le m^\alpha 
n^\mathrm{(0)\alpha}/N_\mathrm{vec}$, where $n^{(0)\alpha}$ is the
particle density of phase $\alpha$, and $k$ numbers each of the
$N_\mathrm{vec}=24$ velocity vectors. In addition, we used periodic
boundary conditions in all three dimensions. We determined that
setting $n^\mathrm{(0)\alpha}>0.6$ for both species guaranteed
immiscibility. In all the simulations we present here we set oil:water
mass fractions to 1:1, specifically at $n^\mathrm{(0)\alpha}=0.7$ for
$\alpha=\mathrm{r,b}$. 

It is experimentally known that the addition of amphiphile into an
immiscible fluid mixture reduces the interfacial tension, as has also
been reported for various lattice gas models in two and three
dimensions \cite{EMERTON97,LOVE01}. To confirm that our model
reproduces this important property, we ran simulations on a
$4\times4\times128$ lattice of a planar interface with surfactant
adsorbed onto it and whose initial density was varied between
simulation runs. The surface tension was calculated with the line
integral along the normal to the interface
\cite{ROWLINSON} 
\be
  \sigma 
  = 
  \int_{-\infty}^{+\infty} \Big[ P_{zz}(z) -
    P_{xx}(z) \Big] \mathrm{d}z            \,,\label{SURFACETENSION}
\ee
\noindent where, for the pressure tensor ${\mathsf
P}\equiv\{P_{ij}\}$, we used the expression \cite{P2}  
\begin{eqnarray}
	{\mathsf P}({\mathbf x})
	&=&
	\sum_{\alpha}\sum_k\rho_k^{\alpha}({\mathbf x}) {\mathbf c}_k
	{\mathbf c}_k 				 	\nonumber\\
	&+&
	\frac{1}{4}\sum_{\alpha,\bar\alpha} g_{\alpha\bar\alpha}
	\sum_{{\mathbf x}'} \Big[ \psi^{\alpha}({\mathbf x})
	\psi^{\bar\alpha}({\mathbf x}') + \psi^{\bar\alpha}({\mathbf
	x}) \psi^{\alpha}({\mathbf x}') \Big]
	(\mathbf{x-x'})(\mathbf{x-x'})				\,. 
	\label{PRESSURETENSOR}
\end{eqnarray}
\noindent We restrict ourselves in this study to nearest-neighbour
interactions, ${\mathbf x}^\prime\equiv{\mathbf x}+{\mathbf c}_k$, and
transversal symmetry allows the second summand within the
integrand in Eq.~(\ref{SURFACETENSION}), which in general is
$\frac{1}{2}(P_{xx}+P_{yy})$, to be simplified as
shown. Equation~(\ref{PRESSURETENSOR}) contains a kinetic (first)
term, the momentum flux, due to the free streaming of particles
corresponding to an ideal gas contribution, plus a potential or virial
(second) term due to the inter-particle momentum transfer derived from
the force (\ref{LBE-FORCE}) \cite{FERZIGER,ALLEN-TILDESLEY}. 

\begin{figure}[htb]
\begin{center}
\includegraphics[angle=0,width=10cm]{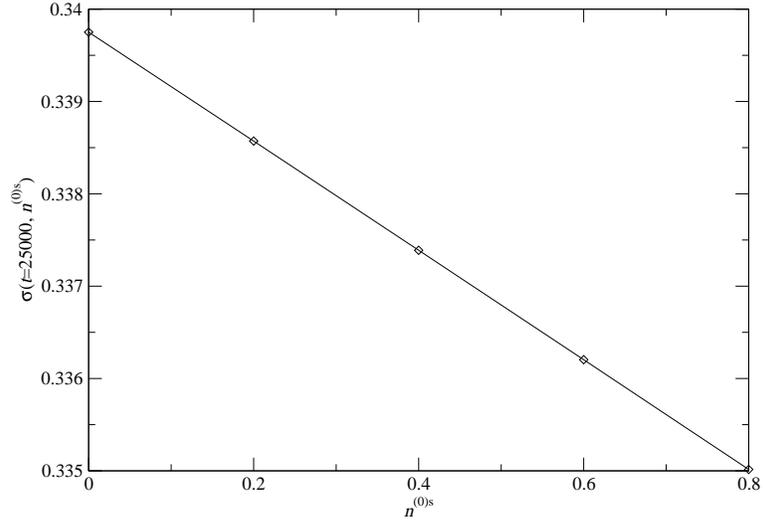}
\caption{Surface tension dependence on the surfactant concentration
(mass fraction, {\em cf.}~Table~\ref{SURF_DENS}) as measured at a
planar interface making use of Eq.~(\ref{SURFACETENSION}). A lattice
of size \mbox{$4\times 4\times 128$} was allowed to evolve up to time
step 25000, and pressure tensor components were measured every 1000
time steps. The surface tension tends to grow with time and reaches a
horizontal asymptote; at that time step the surface tension only
differs in 16\% from that at the previous measurement. Interpolation
serves as a reference to the eye. Coupling constants used were
$g_{\mathrm{br}}=0.08$, $g_{\mathrm{bs}}=-0.006$, and
$g_{\mathrm{ss}}=-0.003$. Oil and water densities used were
$n^\mathrm{(0)R}=n^\mathrm{(0)B}=0.7$. All quantities are reported in
lattice units.}       
\label{SIGMA_VS_SURF}
\end{center}
\end{figure}

Figure \ref{SIGMA_VS_SURF} shows the surface tension $\sigma$ plotted
against initial surfactant density, and details on parameters and
densities used are included in the caption. Notice that in the regime
the binary fluid is in, and for the values of surfactant density we
use, the surface tension decreases linearly with surfactant
concentration. It is entirely possible that there may be departures from
linearity were we to increase the surfactant concentration beyond that
shown in Fig.~\ref{SIGMA_VS_SURF} because of interfacial saturation
with surfactant, as observed in two and three-dimensional lattice-gas
studies \cite{EMERTON97,LOVE01}.

\section{Domain growth kinetics}

We ran simulations starting with a homogeneous mixture of oil and
water particles mixture to which surfactant was randomly added across
on the lattice. Lattice sizes employed were $64^3$ and $128^3$ to
assess finite size effects. Each lattice site was populated with a
density uniformly distributed in the range zero up to the values
summarised in Table \ref{SURF_DENS}.

\begin{table}[!hbp]
\begin{center}
\begin{tabular}{|c|c|c|c|c|c|c|c|c|}
\hline 
simulation run &{\tt 01}&{\tt 02}&{\tt 03}&{\tt 04}&{\tt 05}&{\tt
06}&{\tt 07}&{\tt 08}            			\\
\hline                                            
$n^\mathrm{(0)s}$ & 0.0 & 0.15 & 0.22 & 0.30 & 0.35 & 0.40 & 0.60 & 0.90
\\ 
\hline
$x^\mathrm{s}$ & 0.0 & 0.21 & 0.31 & 0.43 & 0.50 & 0.57 & 0.86 & 1.3 \\  
\hline
\end{tabular}
\end{center}
\caption{\small Surfactant densities employed in the study of the
algebraic-to-logarithmic and logarithmic-to-stretched exponential
transitions. The mass fraction, $x^\mathrm{s}$, is the ratio of
$n^\mathrm{(0)s}$ to $n^\mathrm{(0)b}=n^\mathrm{(0)r}=0.7$, and the
rest of the parameters used were $g_\mathrm{br}=0.08$,
$g_\mathrm{bs}=-0.006$, $g_\mathrm{ss}=-0.003$, masses and relaxation
times set to 1.0, and $\beta=10.0$. The lattice used was sized $64^3$
for all simulation runs except 01, for which it was $128^3$ in order
to avoid finite size effects entering at about $L\approx25$.}
\label{SURF_DENS}
\end{table}

The average size of the oil-water domains is a natural measure of the 
degree of segregation within the mixture. We define it as the inverse first
moment of the spherically averaged oil-water structure function,
$L(t) \equiv 2\pi/k_1(t)$, where 
$k_1(t)\equiv\sum_k kS(k,t)/\sum_k S(k,t)$. The spherically
averaged oil-water structure function, $S(k,t)$, is 
$\sum_\mathbf{\hat k} S({\mathbf k},t)/\sum_\mathbf{\hat k} 1$, 
where the $S({\mathbf k},t)$ is the oil-water structure function, 
\begin{equation}
	S({\mathbf k},t)
	\equiv
	\frac{\varsigma}{V}
	\Big|
		\phi_{\mathbf k}^\prime(t)
	\Big|^2					\,,\label{STRUCT_FUNCT}
\end{equation}
\noindent and $\sum_\mathbf{\hat k}$ denotes summation over the
set of wavevectors contained in the spherical shell
$n-\frac{1}{2}\le \frac{V^{1/3}}{2\pi}k\le n+\frac{1}{2}$, for integer
$n$. Equation (\ref{STRUCT_FUNCT}) is the Fourier transform of the 
spatial auto-correlation function for the oil-water order parameter 
$\phi\equiv\rho^\mathrm{r}-\rho^\mathrm{b}$, where $V$ is the lattice
volume, $\varsigma$ the volume of the lattice unit cell, and 
$\phi_{\mathbf k}^\prime(t)$ is the Fourier transform of the fluctuations 
of the order parameter, $\phi$. Our choice of the structure function,
rather than alternative measures of domain
size such as the 
auto-correlation function, was made on the basis that it is directly
proportional to X-ray or neutron scattering intensities, hence 
facilitating direct comparison with empirical data \cite{GUNTON}. 

\begin{figure}[htb]
\begin{center}
\includegraphics[angle=0,width=12cm]{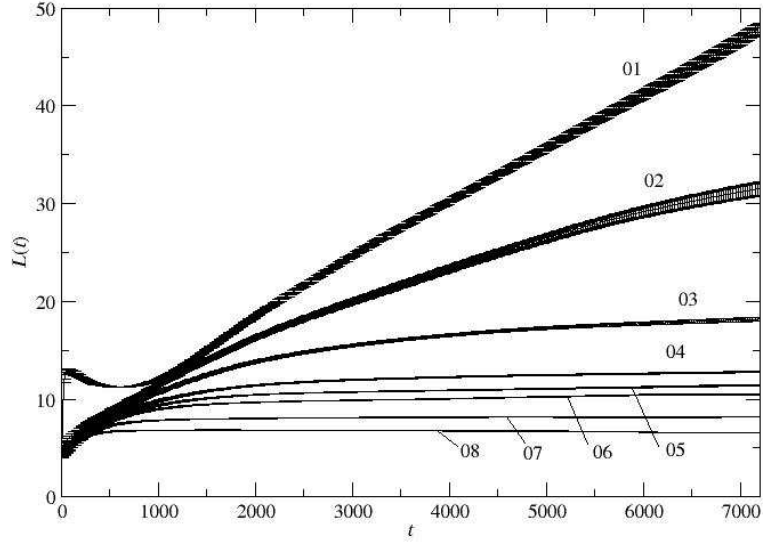}
\caption{Temporal evolution of the average fluid-fluid domain size for
  surfactant concentrations 0.0, 0.15, 0.22, 0.30, 0.35, 0.40, 0.60
  and 0.90 for curves from top to bottom and corresponding to
  simulation runs 01, 02, 03, 04, 05, 06, 07 and 08, respectively
  ({\em cf.} Table \ref{SURF_DENS}). Measurements have been taken every 25
  time steps, and the plots include error bars, which represent the
  uncertainty (one standard error) transmitted from the standard error
  of the structure function spherical average. We used a lattice of
  size $128^3$ for simulation run 01 and $64^3$ for the remaining
  curves, since finite size effects start to creep in for domain sizes
  larger than $L\approx30$. All quantities are reported in lattice
  units. Note that the surfactant-containing fluids lack the
  zero-growth, linear transient found for simulation run 01
  \cite{KAWAKATSU93}.}       
\label{SIZ}
\end{center}
\end{figure}

In Fig.~\ref{SIZ} we plot the temporal evolution of the average domain
size $L$ for the surfactant concentrations of
Table~\ref{SURF_DENS}. The amount of surfactant needed to slow down
the kinetics of the binary immiscible oil-water mixture (simulation
run 01) is seen to be relatively low. We now need to find the growth
laws that best fit these data. Previous simulation studies, for 1:1
oil-water fluid mixtures with or without
surfactant~\cite{P2,EMERTON97,LOVE01,MAZIAR-ET-AL}, have found
algebraic, logarithmically slow and stretched exponential behaviours, as
follows 
\bea
	&a_1(t-b_1)^{c_1}\,,&                     	\label{ALG}\\
	&a_2(\ln t)^{c_2}\,,&                        	\label{LOG}\\
	&a_3-b_3\exp\Big[-c_3(t-d_3)^{e_3}\Big]\,,&	\label{EXP}
\eea
\noindent to be those characterising the temporal growth of the
average domain size, $L(t)$, of an oil-water mixture without
surfactant, Eq.~(\ref{ALG})~\cite{P2}, and when surfactant is added
above a minimum threshold concentration, Eq.~(\ref{LOG}), and at a
sufficiently high amphiphile concentration, Eq.~(\ref{EXP}), the
latter being a regime for which arrested growth is reached at late
times.  
The coefficients $a_i$ and $b_i$ ($i=1,2,3$) are fitting
parameters. While we shall take these functional forms as suggested
choices, we would also like to find out how closely they in fact fit
our data.

Linearity in the $(t,L)$ data cloud on a log-log plot would permit us
to ascertain whether or not the data follow Eq. (\ref{ALG}),
regardless of the zero-time offset value $b_1$ since this is a
horizontal displacement. To find out which data may be better fit by
Eq. (\ref{LOG}), we would require $(\log t,L)$ pairs of data in a
search for linearity on a log-log plot. This method, however, is not
likely to be of much help given the small difference between plots of
the logarithm of a data series and plots of the logarithm of such
logarithmic data, as we shall see. We therefore prefer to adopt the
criterion of considering candidates for the model of Eq.~(\ref{ALG})
from the log-log linearity method, while resorting to both visual
inspection and a search for a reduced chi-squared statistic
($\chi^2/\mathrm{ndf}$) close to 1.0 in order to identify a slower
growth such as that of Eq.~(\ref{LOG}) ($\mathrm{ndf}$  is the number
of degrees of freedom). Finally, Eq.~(\ref{EXP}) possesses a
distinctive horizontal asymptote which best fits data whose domain
growth at late times is fully arrested.

\begin{figure}[htb]
\begin{center}
\begin{tabular}{r l}
(a) & \includegraphics[angle=0,width=10cm]{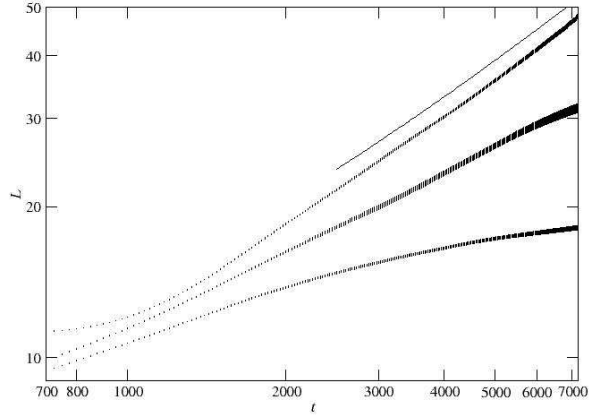}
	\\
    & 	\\
    &  	\\
(b) & \includegraphics[angle=0,width=10cm]{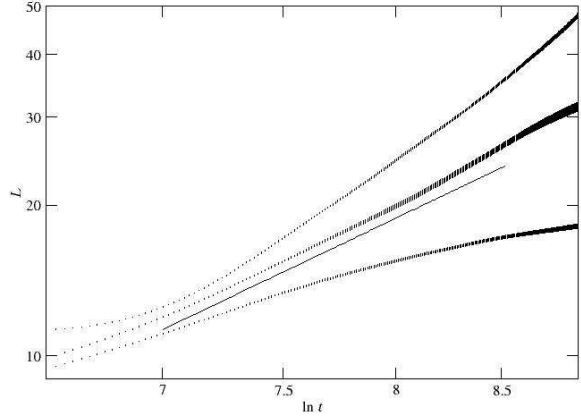}
	\\
\end{tabular}
\caption{Panel (a) shows the time evolution of the average domain size
  for simulation runs 01, 02 and 03, see Fig. \ref{SIZ}. The log-log
  scale helps to visually detect behaviours following
  Eq.~(\ref{ALG})---in this case, that of the uppermost curve. The
  straight line above it serves as a guide to the eye only and its
  slope is given by $c_1$ in Table~\ref{FITS}. Panel (b) shows the
  evolution of 
  the average domain size with the logarithm of the time step for
  simulation runs 01, 02 and 03, on a log-log scale. This is useful to
  discriminate growth between that of Eq.~(\ref{ALG}) and
  Eq.~(\ref{LOG}); see Table~\ref{FITS} for the fitting parameters. The
  straight solid line shown indicates a good fit to Eq.~(\ref{LOG})
  for $t>1100$ ($\ln t\approx7$). For $t<1100$, the curve is better
  fit by Eq.~(\ref{ALG}), albeit still quite poorly. Measurements have
  been taken every 25 time steps, and the plot includes error bars
  representing 
  the uncertainty (one standard deviation) of the spherical averaged
  structure function. We use a $128^3$ lattice for simulation run 01
  and a $64^3$ lattice for the remainder. All quantities are reported in
  lattice units.}     
\label{SIZ_LOG-LOG}
\end{center}
\end{figure}

From the linearity of curves in Fig. \ref{SIZ_LOG-LOG} we can infer
that simulation runs 01 and 02 (Table \ref{SURF_DENS}) are good
candidates for the growth model of Eq. (\ref{ALG}). Figure 
\ref{SIZ_LOG-LOG}, however, leads to the same conclusion, as
expected given the small difference between these two plots. We then 
resort to looking at the $\chi^2/\mathrm{ndf}$ statistic in assessing
how well Eqs. (\ref{ALG}) and (\ref{LOG}) fit simulation
runs 01, 02 and 03, see Table \ref{FITS}. The binary immiscible fluid
simulation run 01, with no surfactant present, exhibits an exponent
consistent with the system being in a crossover between the known
diffusive ($t^{1/3}$) and viscous hydrodynamic ($t^{1.0}$) regimes,
already reported for binary immiscible fluids simulated with the
lattice-BGK model we employ in this paper \cite{P2}. Simulation
run 02 has the peculiarity that Eq. (\ref{ALG}) holds (poorly) only
during an initial transient, and Eq. (\ref{LOG}) takes over to give a
very good fit at
later times, $t>1100$. This transient is due to the time required by
the surfactant to adsorb onto the interface and affect the binary
immiscible interfacial dynamics. Finally, simulation run 03 is best 
fit by Eq.~(\ref{LOG}), although the high $\chi^2/\mathrm{ndf}$ value 
indicates that the data contain more detail than the model does. In
addition, from Fig.~\ref{SIZ_LOG-LOG}b, this mixture segregates at a
slower speed than that given by Eq.~(\ref{LOG}), yet it does not reach
total arrest, at least up to 7200 time steps. Rather, total arrest is seen 
at higher surfactant concentrations, as in runs 06 and 07
(see Fig. \ref{SIZ_OSCILL}). We conclude that simulation run 03
represents a fluid which is in a transition regime between
the logarithmic and the stretched exponential behaviours. A similar
behaviour was previously observed by others using lattice-gas
methods in two \cite{EMERTON97} and three-dimensions \cite{LOVE01},
and lattice-Boltzmann methods in two dimensions
\cite{MAZIAR-ET-AL}. Emerton {\em et al.}, using a two-dimensional
lattice-gas model, reported the divergence of the coefficients of
Eq. (\ref{EXP}) in an attempt to fit data for which total growth
arrest had not been achieved \cite{EMERTON97}. The fits to our data,
which include error bars, also produced the same divergences. Their
fluid mixtures as well as ours, we conclude, were, rather, in a
transitional regime well described by a growth law slower than
Eq. (\ref{LOG}) which still allowed for domain growth. It is, however, 
possible that growth arrest could be achieved at later times; 
this pre-arrest regime would then be a long-lived transient.

\begin{table}[!hbp]
\begin{center}
\begin{tabular}{|c|c c|c c|}
\hline  
simulation run & $c_1$ & $\chi^2/\mathrm{ndf}$ & $c_2$ &
$\chi^2/\mathrm{ndf}$							\\ 
\hline  
{\tt 01}  & $0.896\pm 0.007$ & 0.18 & --- & --- 	 		\\ 
{\tt 02}  & $0.644\pm 0.004$ & 7.5  & $3.850\pm 0.010$ & 0.92 		\\ 
{\tt 03}  & --- & --- & $2.649\pm 0.022$ & 39 				\\  
\hline
\end{tabular}
\end{center}
\caption{\small Fits of the average domain size growth with time to
the models of Eqs.~(\ref{ALG}) and (\ref{LOG}) for simulation runs
corresponding to surfactant mass fractions 0.0, 0.21 and 0.31, from top to
bottom, respectively, as detailed in Table~\ref{SURF_DENS}. Lattice
sizes used were $128^3$ for simulation run 01 and $64^3$ for the
rest. Poor fits are indicated as blank fields. Simulation run 02 shows
two behaviours in its temporal evolution, Eq.~(\ref{ALG}) for $t<1100$
and Eq.~(\ref{LOG}) for $t>1100$. Note the very good value of the
$\chi^2/\mathrm{ndf}$ statistic for the latter. The poor value of the
statistic for simulation run 03 indicates that Eq.~(\ref{LOG}) is
insufficient and a more detailed model is required, albeit not
Eq.~(\ref{EXP}).} 
\label{FITS}
\end{table}

\begin{figure}[htb]
\begin{center}
\includegraphics[angle=0,width=10cm]{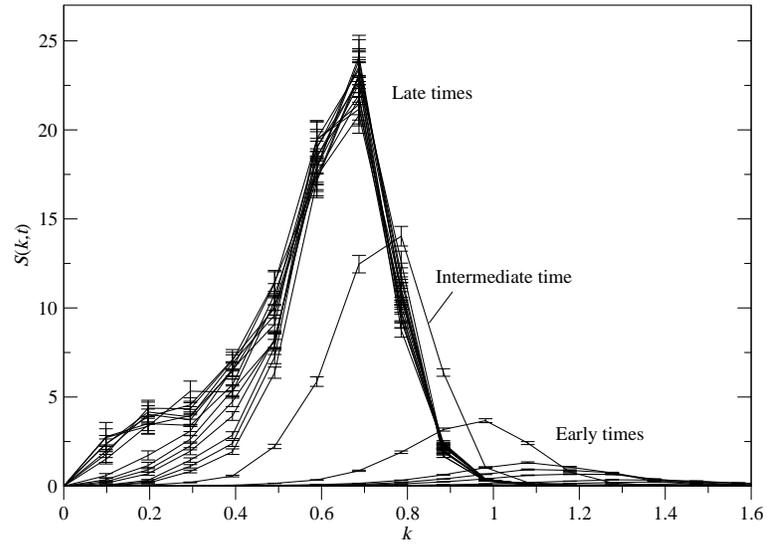}
\caption{Spherically averaged structure functions for the oil-water
  order parameter simulation run 06 ({\em cf.} Table 
\ref{PARAMS_OSCILL}). According to how close to asymptotic 
behaviour the distribution of domain sizes appears to be, we have classified
simulation times for this simulation run in three groups: early times (time
steps 25, 50, 75, 125, 150, 175, and 300 in the plot), intermediate
times (time steps roughly from 800 to 1700), and late times (time
steps 1800, 2300, 2800, 3300, 3800, 4300, 4800, 6000, 10000, 14000,
18000, 22000, 26000, and 30000 in the plot). Error bars represent the
  standard error of the shell average. Lattice size is $64^3$. All
  quantities are reported in lattice units.}  
\label{SF_09}
\end{center}
\end{figure}

We now look at wavenumbers of the spherically averaged structure 
function, $S(k)$, other than the first moment, already provided by
$L(t)$. Figure \ref{SF_09} shows the spherically averaged structure 
function for simulation run 06 at several time steps. The temporal
evolution of the curves 
 
resembles the segregation kinetics for binary immiscible fluid mixtures, 
except that domain growth arrest for late times makes them tend to 
superimpose. Note that a hump appears at these times, indicating the 
formation of structures, statistically weak, of size close to half the 
lattice side length. Inspection of $\phi({\mathrm x})$ snapshots
suggests the spurious presence of elongated domains of such sizes which 
are extended rather than folded. At the late times we examined, these 
elongations tend to vanish or fold. Still in Fig. \ref{SF_09}, it is
worth noting that for all length scales above a threshold (about
$k<0.9$), curve superposition is not sharp. This is a consequence of
the fact that for the fluid 
composition of simulation run 06, and those of higher surfactant
concentrations, there are small temporal oscillations in $S(k)$. All
these mixtures have in common that they have achieved total growth
arrest---in fact, $L(t)$ decreases in time for simulation run~08, as
we shall see later on and discuss in more detail. Oscillations in the
structure function and a hump at low wavenumbers have been 
reported previously in a hydrodynamic Langevin model of sponge phase
dynamics, using field-theoretic methods \cite{GOMPPER}. However, this
approach did not consider the amphiphile concentration explicitly but,
rather, embedded it into a Ginzburg-Landau free energy through the
surface tension, in a scenario where amphiphile relaxation is assumed
to be fast compared to that of the oil-water order parameter. 

\begin{figure}[htb]
\begin{center}
\begin{tabular}{r l}
(a) & \includegraphics[angle=0,width=10cm]{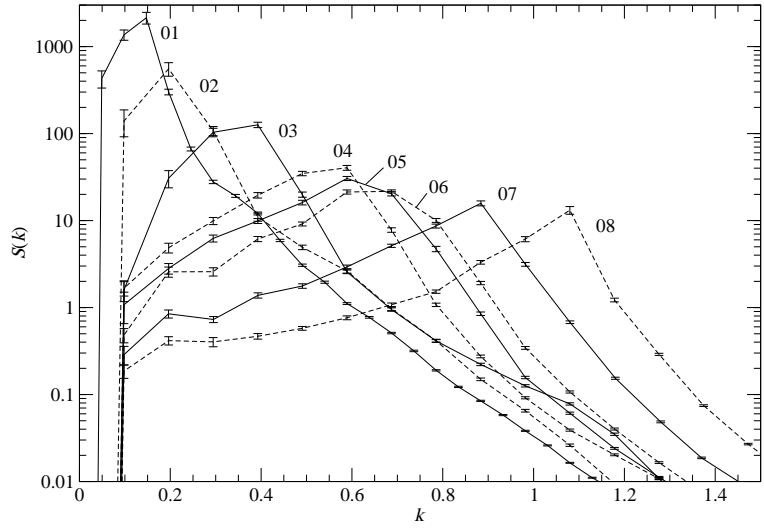} 
\\
 & \\
 & \\
 & \\
(b) & \includegraphics[angle=0,width=10cm]{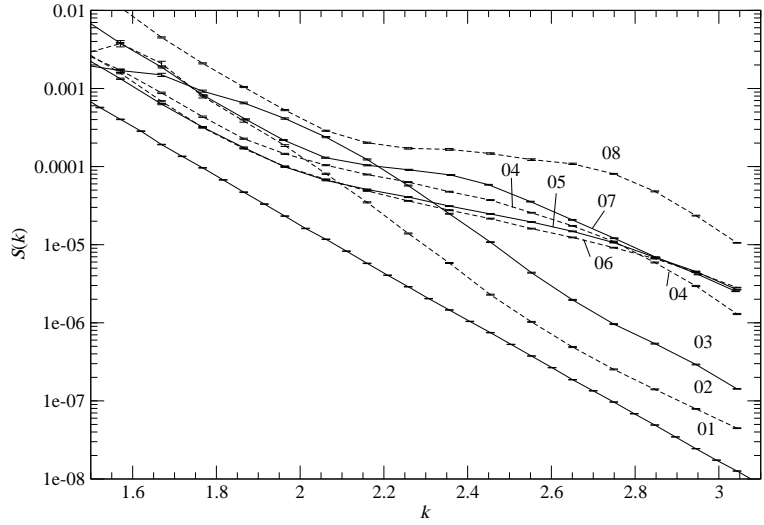} 
\end{tabular}
\caption{Log-linear plots of the spherically averaged structure
  functions at time step 7200 for increasing surfactant concentrations
  indicated by the numerical labelling on each curve, corresponding to
  simulation runs 01, 02, 03, 04, 05, 06, 07 and 08, respectively,
  {\em cf.} Table~\ref{SURF_DENS}. For large wavelengths, panel 
(a), we can see how the peaks move to higher wavenumbers, decrease in
height, and broaden. Note that for short wavelengths, panel (b), 
the only straight tail is for curve $n^\mathrm{(0)s}=0.0$, whose slope
is $-4.46\times10^{-4}$. Error bars represent one standard error of
the shell average $S(k)$. Lattice size is $128^3$ for simulation run
01 and $64^3$ for the others. All quantities are reported in lattice
units.}
\label{SF_VARY_SURF}
\end{center}
\end{figure}

In Figs. \ref{SF_VARY_SURF}a and \ref{SF_VARY_SURF}b we show the
spherically averaged structure function at time step 7200 of the
mixtures in Table \ref{SURF_DENS}, for the larger and smaller length
scales, respectively. As the initial density of surfactant is
increased in a series of replica initially homogeneous
water-oil-surfactant fluid mixtures, as indicated by Table
\ref{SURF_DENS}, it is expected that the oil-water structure function
peaks will move to higher wavenumbers, decrease in intensity, and
broaden \cite{GOMPPER&SCHICK,GOMPPER&SCHICK2,GOMPPER&SCHICK3}. This is
indeed what we observe in Fig. \ref{SF_VARY_SURF}a. Note that at
smaller length scales, Fig. \ref{SF_VARY_SURF}b, the exponential decay
of the structure function that occurs for simulation run 01 does not
hold for the ternary amphiphilic mixtures. This can be explained by
the contribution of small micellar structures that form in the
bulk of each immiscible phase, more likely to take place for mixtures
of higher surfactant concentration. Indeed, in
Fig.~\ref{SF_VARY_SURF}b, the latter exhibit the most manifest 
deviations.

\section{Self-sustained oscillations}
\label{OSCILL}

Arrest of domain growth occurs for high surfactant density only,
{\em cf.} Fig.~\ref{SIZ}, as expected. Further inspection, however, shows
that not only are there small temporal oscillations of the average
domain size, as we mentioned at the end of the last section, but also
that they do not die out during the simulation window. Similarly to
what was previously reported
using a bottom-up lattice-Boltzmann method in two dimensions akin to
the one employed here \cite{MAZIAR-ET-AL}, the amplitude of the
oscillations is very small compared to the average domain size, and
smaller than previous lattice-gas simulations in two and three
dimensions \cite{EMERTON97,LOVE01}. The fact that, in lattice-gas
methods, these oscillations persist after ensemble averaging is
consistent with their occurrence in lattice-Boltzmann approaches,
since the latter are effectively ensemble-averaged versions of the
former. Since the systems we simulate are dissipative and isolated
(there is no mass or momentum exchange with external sources),
oscillations, however, are expected to die out at sufficiently late
times. 

Motivated by the observation of oscillating average domain sizes, we
performed additional simulation runs in order to check the role of the
coupling constants $g_\mathrm{ss}$ and $g_\mathrm{bs}$ in the
reproduction of such oscillations, our hypothesis being that both an
increased surfactant-surfactant interaction and an increased tendency
for surfactant to adsorb on the interface might be expected to have an
influence on their frequency and amplitude. In Table
\ref{PARAMS_OSCILL} we summarise the parameters used in the new
simulation runs (09 and 10) along with those of previous `oscillating
fluid mixtures.'

\begin{table}[!hbp]
\begin{center}
\begin{tabular}{|c|c|c|c|c|}
\hline  
simulation run 	& $n^\mathrm{(0)s}$ & $x^\mathrm{s}$ & $g_\mathrm{ss}$
& $g_\mathrm{bs}$					\\ 
\hline  
{\tt 06}  &  0.40   & 0.57 &	-0.0030	& -0.006 \\ 
{\tt 07}  &  0.60   & 0.86 &	-0.0030	& -0.006 \\ 
{\tt 08}  &  0.90   & 1.3  &	-0.0030	& -0.006 \\ 
{\tt 09}  &  0.60   & 0.86 &	-0.0045	& -0.006 \\ 
{\tt 10}  &  0.60   & 0.86 &	-0.0030	& -0.009 \\ 
\hline
\end{tabular}
\end{center}
\caption{\small Parameters employed in studying domain size
oscillations, whose onset occurs for surfactant mass fractions
$x^\mathrm{s}\ge0.57$; the remaining parameters of the model are
stated in the caption of Table~\ref{SURF_DENS}, also for the
additional runs 09 and 10. In lattice units.}  
\label{PARAMS_OSCILL}
\end{table}

\begin{figure}[htb]
\begin{center}
\includegraphics[angle=0,width=14cm]{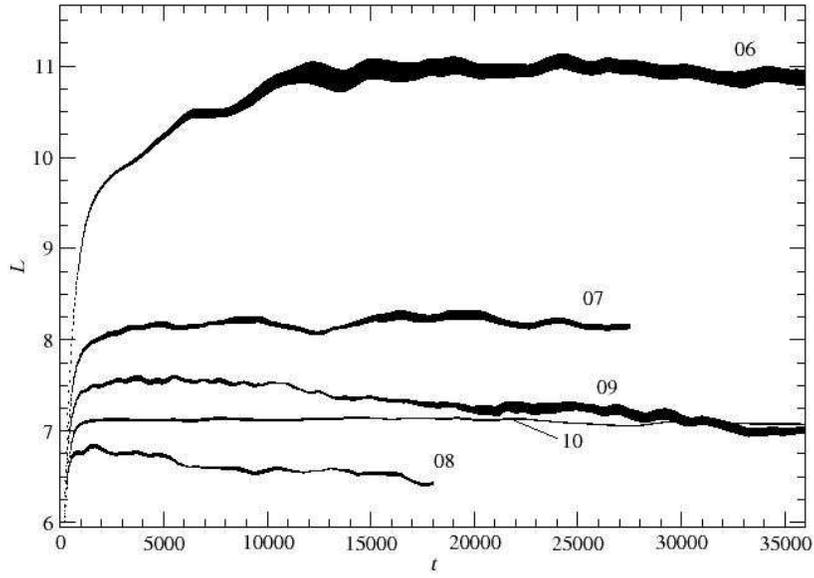}
\caption{Temporal evolution of the average domain size for simulation
runs 06, 07, 09, 10, and 08, as seen from top to bottom at $t=10000$
({\em cf.} Table~\ref{PARAMS_OSCILL}). Measurements have been taken every 25
time steps; error bars are included and represent the uncertainty
(``one sigma") transmitted from the standard error of the spherically
averaged structure function. {\em Caveat lector}: an oscillation in
the average domain size is genuinely representative of oscillations in
the domain sizes only if error bars are smaller than the oscillation
amplitude. Lattice size is $64^3$. All quantities are reported in
lattice units.}
\label{SIZ_OSCILL}
\end{center}
\end{figure}

\begin{figure}[htb]
\begin{center}
\includegraphics[angle=0,width=10cm]{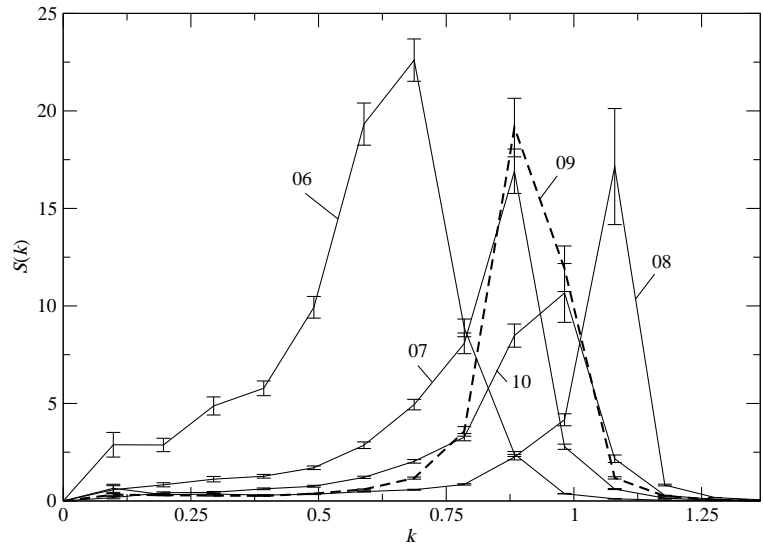}
\caption{Structure functions at late time step 17000 for
simulation runs 06, 07, 09, 10 and 08 ({\em cf.} Table \ref{PARAMS_OSCILL}).
Error bars represent the standard error of the shell average
$S(k)$. Lattice size is $64^3$. All quantities are reported in lattice
units.} 
\label{SF_ZOOMED-IN_OSCILL}
\end{center}
\end{figure}

Figure~\ref{SIZ_OSCILL} shows the temporal oscillations in the average
domain size for the mixtures of Table \ref{PARAMS_OSCILL}, and
Fig.~\ref{SF_ZOOMED-IN_OSCILL} shows their structure functions at time
step 17000. All these mixtures exhibit domain growth arrest; interestingly,
Fig.~\ref{SIZ_OSCILL} shows that the average domain size shrinks in time
for some of them (mixtures 08 and 09). In addition, we uncover the
role that the coupling constants $g_\mathrm{ss}$ and $g_\mathrm{bs}$
have in the oscillations: whilst increasing $|g_\mathrm{ss}|$ seems to
enhance their frequency, an increase in $|g_\mathrm{bs}|$ drastically
dampens them and reduces their amplitude. However damped the
oscillations of simulation run 10 may seem, zooming into smaller
scales reveals the existence of minute oscillations (less than 0.10
lattice sites in amplitude), which is not the case for simulation runs
01 through to 05. (Note that the length scales reported in
Fig.~\ref{SIZ_OSCILL} are lattice averages; an amplitude being less
than one lattice site hence remains physically meaningful.)
Oscillations are, therefore, the signature of all growth-halted
regimes. 

The structure function plots of Fig. \ref{SF_ZOOMED-IN_OSCILL} provide
further insight into the role of coupling constants
$g_\mathrm{ss}$ and $g_\mathrm{bs}$ in the oscillation dynamics. Note
that mixture 08 produces a peak of intensity similar to that of
mixture 07, a feature already seen at much earlier times (see
Fig. \ref{SF_VARY_SURF}a). This peak height similarity could have been
ascribed to a transient, such as turned out to be the case for the
difference in peak intensities between mixtures 06 and 07; however, it
persisted in time. Mixture 09 also shows a peak intensity similar to
that of mixture 07. Peak intensities bear a direct relation to the
steepness of oil-water domain walls and, hence, to their surface
tension. The fact that increasing the surfactant concentration (in
mixture~08 compared to mixture~07) does not reduce the surface
tension denotes that the interface is close to its saturation limit
with respect to
surfactant adsorption. If enough surfactant is dispersed in the bulk,
a process of diffusion towards and adsorption onto the interface could
continue to occur, much slower compared to the initial adsorption
leading to growth arrest, which could explain the slow domain size
reduction. In the cases of simulation runs 08 and 09, close to
interface saturation, surfactant concentration in the bulk is high. An
amphiphilic mixture being close to the saturation limit implies that
the value of its surface tension is the lowest among all amphiphilic
mixtures sharing the same composition, relaxation times and coupling
constants $g_{\alpha\bar{\alpha}}$. Surface tension may be further
reduced only by allowing more surfactant molecules onto the interface,
which can be done by increasing $|g_\mathrm{bs}|$. This is exactly
what we observe in Fig.~\ref{SF_ZOOMED-IN_OSCILL} for fluid
composition 10.

\begin{figure}[htb]
\begin{center}
\begin{tabular}{r l}
(a) & \includegraphics[angle=0,width=11.5cm]{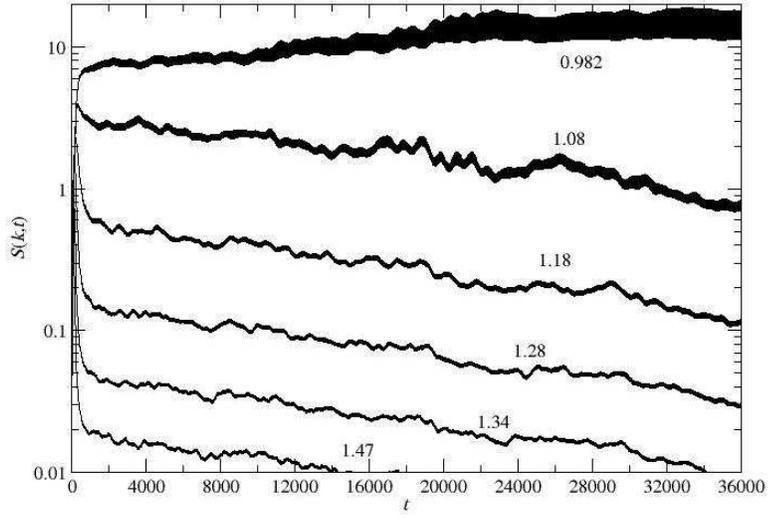}
\\
 & \\
 & \\
 & \\
(b) & \includegraphics[angle=0,width=12.5cm]{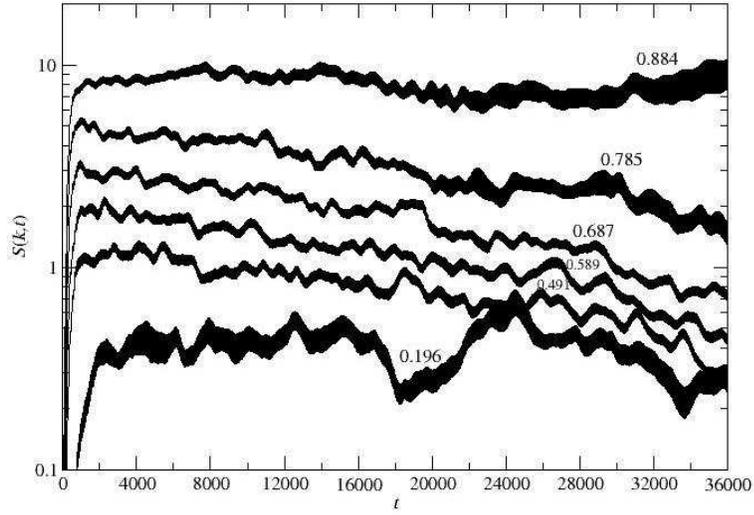}
\end{tabular}
\caption{Temporal dependence of the structure function for simulation
run 09, {\em cf.} Table \ref{PARAMS_OSCILL}. Panels (a) and (b) show
the short and long wavelengths, respectively. Measurements have been
taken every 25 time steps; error bars have been included. Lattice
size is $64^3$. All quantities are reported in lattice units.}         
\label{SF_vs_t_15}
\end{center}
\end{figure}

As we saw in Fig.~\ref{SF_09}, small oscillations in the average
domain size indicate that the structure function varies in time back
and forth between distributions of sizes which are close to each
other. The first moment of such distributions, as studied in
Fig.~\ref{SIZ_OSCILL}, may not be representative of the dynamics at
other length scales, as we shall see immediately. In
Fig.~\ref{SF_vs_t_15} we show the temporal evolution of $S(k)$ for
mixture 09 for a range of wavelengths. Note three characteristic
features of the $S(k)$ curves: they all oscillate, decrease for
$k<0.785$ and $k>1.08$, and increase or remain stationary in the
long time average for $k\approx0.884$ and $k\approx0.982$. This
behaviour corresponds to the sharpening of the distribution $S(k)$
with time. Modes with $k>1.28$ ($L<4.91$) decay fast enough
($S(k)<0.1$ for $t\approx1000$) for them to be negligible in terms of
their contribution to the fluid mesostructure. Other decreasing modes
take much longer ($t>30000$) to vanish. 

\begin{figure}[htb]
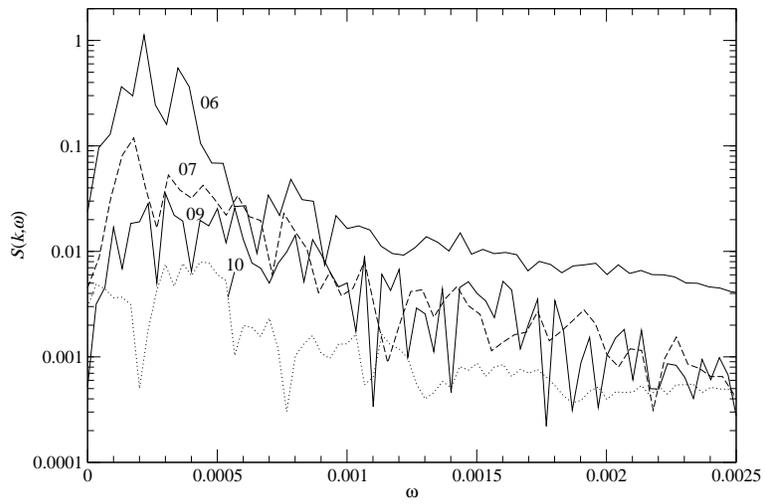

\begin{center}
\begin{tabular}{r l}
(a) & \includegraphics[angle=0,width=10cm]{eps/other09a.eps}
\\
 & \\
 & \\
 & \\
(b) & \includegraphics[angle=0,width=10cm]{eps/other09b.eps}
\end{tabular}
\caption{Frequency power spectra of the structure function for
simulation runs 06, 07, 09 and 10 ({\em cf.} Table \ref{PARAMS_OSCILL}) at
wavenumber 0.589. Panel~(b) is a blowup of panel~(a) for
long periods of oscillation. Error bars were neither considered for the
Fourier analysis nor plotted here. All quantities are reported in 
lattice units, and $\omega$ is the inverse period.}        
\label{SFFT_09.12.15.16_ALL-FREQS}
\end{center}
\end{figure}

Our study of the oscillations would be incomplete without looking at
frequency power spectra. The time series we analyse correspond to
$S(k=0.589)$ of fluids 06, 07, 09 and 10; this choice is made on the
basis that this wavenumber apprehends characteristic features of each
data set. From each time series we subtracted its longest waves
(i.e. its envelope), computed as the average
$\frac{1}{\lambda}\sum_{t^\prime}S(k,t^\prime)$, $\lambda$ being a lag 
large enough so as to decouple high-frequency from low-frequency waves
($\lambda=5000$ time steps), the sum extending over the interval
$t-\lambda/2 \le t^\prime \le t-\lambda/2$.  The Fourier transform of
the resulting time series we take as the definition of $S(k,\omega)$,
{\em cf.} Fig. \ref{SFFT_09.12.15.16_ALL-FREQS}. Note therein two high peaks
for simulation run 06, and a collection of weak peaks (which we define
as those whose heights are less than 5\% the height of the largest
peak) occuring for 
higher frequencies. An increase in surfactant density (simulation run
07) causes the number of excited high-frequency modes to grow
slightly, yet they also decrease in intensity. Simulation run 09,
which differs from mixture 07 in having an increased
$|g_\mathrm{ss}|$, very clearly exhibits a substantial increment in
the number of excited high-frequency modes. Finally, the spectrum for
mixture 10 corroborates the quenching effect on fluctuations caused by
increasing  $|g_\mathrm{bs}|$.

\begin{figure}[htb]
\begin{center}
\begin{tabular}{r l}
(a) & \includegraphics[angle=0,width=7cm]{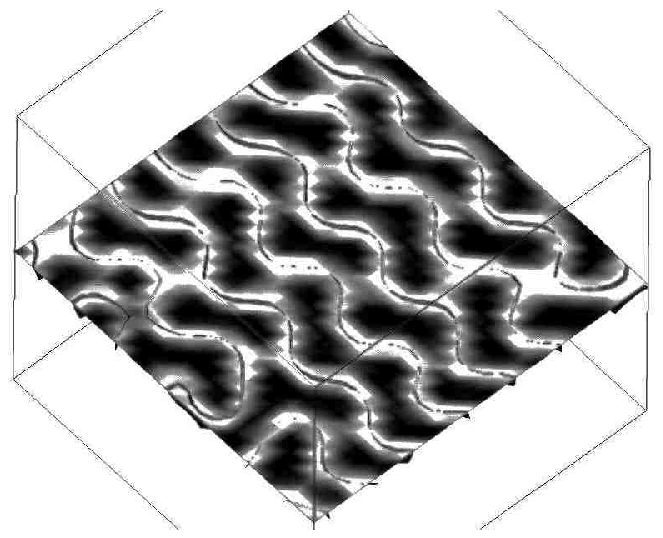}
\\
\\
(b) & \includegraphics[angle=0,width=7cm]{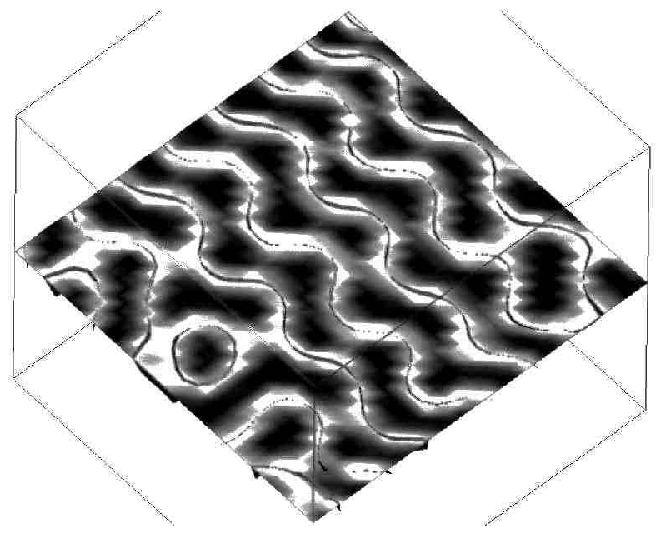}
\\
\\
(c) & \includegraphics[angle=0,width=7cm]{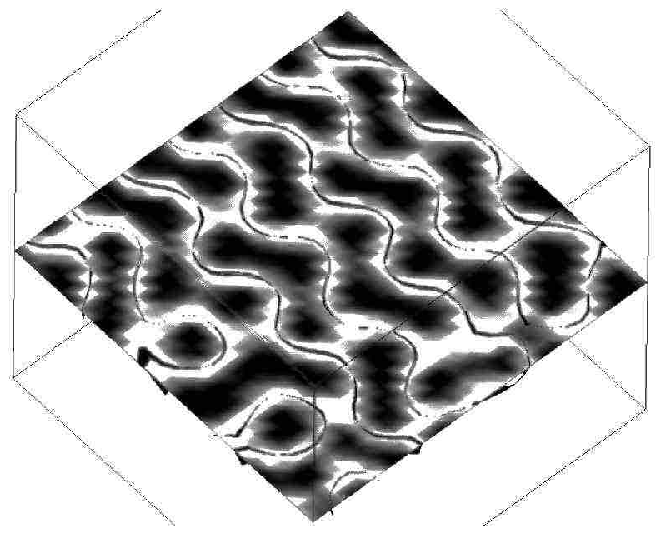}
\end{tabular}
\caption{Slices of the surfactant density in $33^3$ subsets of a
$128^3$ lattice, for composition 09 ({\em cf.~}Table
  \ref{SURF_DENS}). Panels (a), (b) and (c) are snapshots at 
  $t=14000$, 14600 and 15000 time steps, respectively, which are times
  for which the structure is close to equilibration. We only show
  regions where surfactant density is the highest
  ($0.31\le\rho^\mathrm{s}\le0.35$), in grey and white, where lighter 
  shading denotes a higher value. We can see that the surfactant
  mainly concentrates around the oil-water interface
  ($\phi(\mathbf{x})=0$), whose intersection with the slice is
  depicted as open undulating or closed lines. Also, there are
  ordered, crystalline regions along with smaller regions lacking
  long-range order and evolving in time. Finally, $\rho^\mathrm{s}$ is
  non-uniform on the interface (as regions with lighter shading show),
  favouring the formation of ``surfactant bridges'' between adjacent
  portions of the interface; this leads to Marangoni effects which
  account for the observed oscillatory behaviour. All quantities are
  in lattice units.}      
\label{SUR_COLOUR}
\end{center}
\end{figure}

The term Marangoni instability describes a convective flow caused
wherever an inhomogeneous temperature or mass distribution locally
alters the interfacial tension \cite{KOVALCHUK}. By visualising the
oil-water interface for mixtures 06 to 10 we observed that the density
of adsorbed surfactant is not evenly distributed on it; hence the
conditions are set for the appearance of Marangoni instabilities. 
Figure~\ref{SUR_COLOUR} displays the late time evolution of a
subdomain of a fluid of the same composition as simulation run 09 but 
simulated on a larger ($128^3$) lattice. We display the surfactant
density on a slice through 
the mid-plane of the subdomain, along with the locus of the oil-water
interface depicted as an isosurface cropped close to the plane. 
Surfactant inhomogeneities on the interface are evident from these
images, as well as the existence of a slow, creeping flow. Distinctive
features include the regularity of the order parameter (which we shall
study in detail shortly), the existence of high surfactant density
necks bridging adjacent portions of the interface, and local regions
where regularity is absent, reminiscent of the defects in crystalline
materials, which possess their own larger-scale dynamics.

\section{The sponge and gyroid mesophases}

\begin{figure}[htb]
\begin{center}
\includegraphics[angle=0,width=17cm]{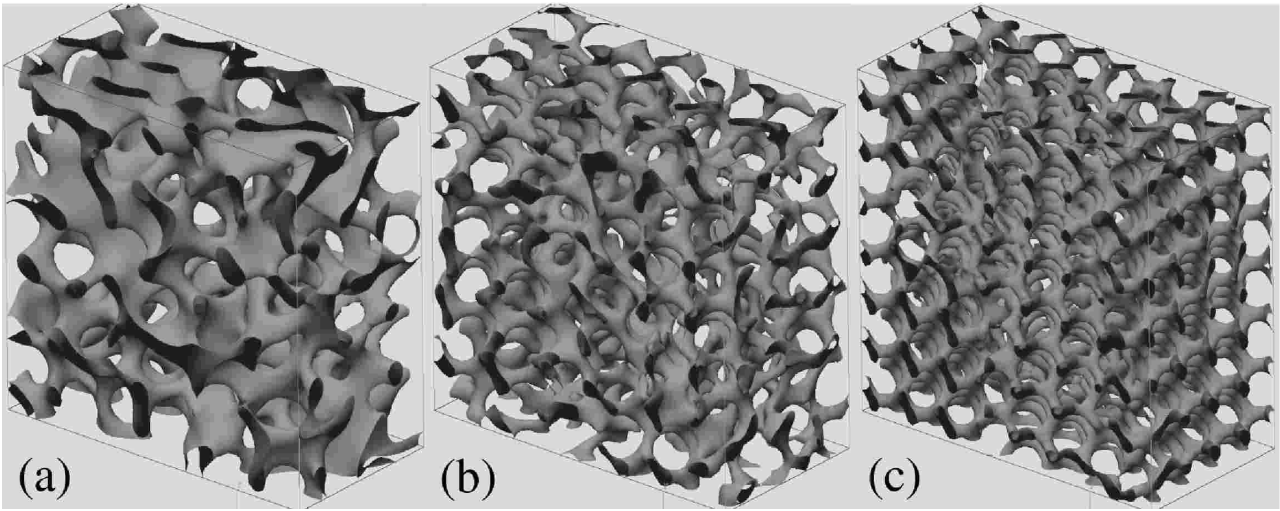}
\caption{Equilibrium minority-phase order parameter isosurfaces
$\phi(\mathbf{x})=0.37$ (in lattice units), taken at time step
$t=30000$, at which $-0.69\le\phi\le 0.68$ over the whole
lattice. Snapshot (a) corresponds to simulation run~06, snapshot (b)
to simulation run~07, and snapshot (c) to simulation run~08, {\em
cf.}~Table~\ref{SURF_DENS}. Shown are $16\le z\le 48$ slabs of a  
$64^3$ lattice. Note that increasing surfactant concentration leads to
an increased ordering in the mesostructure: simulation run~06
exhibits sponge-like features, whereas simulation run~08 is a liquid
crystalline cubic gyroid mesophase. Snapshot (b) is a crossover
state in this lyotropic transition---a ``molten gyroid".}     
\label{ISOSURF_06.07.08}
\end{center}
\end{figure}

It is known that, in an initially homogeneous 1:1 oil-water mixture,
the arrest of phase segregation experienced through the addition of
sufficient amphiphile can lead to the formation of a thermodynamically
stable bicontinuous
sponge phase \cite{LOVE01,EMERTON97,KAWAKATSU93}. In
Fig.~\ref{ISOSURF_06.07.08} we show the late time morphologies for
fluid compositions 06, 07 and 08. They are displayed as the  
$\phi(\mathbf{x})=0.37$ isosurface, corresponding to a water-in-oil,
``rod-like" scenario, where water is a minority phase and oil is in
excess (the order parameter ranges as $-0.69\le\phi\le 0.68$ over the
lattice at that time slice). The structure suggested by minority-phase
isosurfaces and the structure of the interface ($\phi(\mathbf{x})=0$)
for fluid composition 06 resembles that of a microemulsion, for
which structural disorder is the predominant feature. Fluid
composition 08, by contrast, shows an evident resemblance to
minority-phase images seen in transmission electron microtomography of
the gyroid ``G" cubic phase \cite{LAURER}. The morphology is an
interweaving, chirally symmetric, three-fold coordinated, bicontinuous
lattice. Fluid mixture 07 seems to be a crossover, {\em conatus}
structure, sharing a substantial amount of disorder with the presence
of three-fold coordinated ``unit cells"; the latter can be seen as
vestigial in fluid system 06. Fluid systems 09 and 10 show that this
sponge$\leftrightarrow$gyroid structural order-disorder
transition not only occurs via an increase in surfactant concentration
(a lyotropic transition), but in the interaction strength between
surfactant with itself and with the interface. We leave for further
work a systematic investigation of the
$\{n^\mathrm{(0)s},\,g_\mathrm{bs},\,g_\mathrm{ss},\,g_\mathrm{br}\}$ 
parameter space in mapping out the equilibrium mesostructures' phase
diagram. In this endeavour, recently developed {\em compusteering} tools
\cite{COMPUSTEERING,ReG} may prove valuable in optimising expensive
simulation time: they allow the user to postprocess and visualise the
compute job's output at run-time with negligible turnaround times, and
eventually temporarily stop execution in order to modify simulation
parameters which are fed back into the algorithm on immediate
restart. 

Finite size effects can play an important role in the stabilisation of
fluidic structures like these, given that we are using periodic
boundary conditions. With this in mind, and using the same parameters
as for mixture 09, we computed the wavenumber-averaged difference
$\langle\Delta_{N,N^\prime}S\rangle$ for each time step of evolution
of the spherically averaged structure function $S(k)$ between lattices
of sizes $N^3$ and $N^{\prime 3}$, where $N,N^\prime=64,128,256$. Note
that the lattice size is increased eight and sixty-four times from the
original $64^3$ size. Finite size effects would be present if
$\langle\Delta_{N,N^\prime}S\rangle$ were larger than the error
derived from the differences and the averages. Nonetheless we found
$\langle\Delta_{N,N^\prime}S\rangle$ to be larger than the error (27\%
larger on average for $N=128$ and $N^\prime=256$), the fact that it
strongly decreased with $N$
(i.e.
$\langle\Delta_{128,256}S\rangle\approx0.38\langle\Delta_{64,128}S\rangle$) 
provides the confidence necessary to assert that finite size effects
are not significant in the $N=128$ simulations we are about to
report. Moreover, as we shall see immediately,  
since the structures corresponding to  a $64^3$ lattice exhibit the
same morphologies as do the $128^3$ and $256^3$ cases, the qualitative 
features of the former are the same as those for the asymptotic limit
$N\to\infty$. This also extends to the oscillation of the
structure function: the equivalent of Fig.~\ref{SF_vs_t_15} for
the $128^3$ and $256^3$ cases (not shown) exhibits similar features,
albeit including more wavenumbers for which $S(k)$ grows.

\begin{figure}[htb]
\begin{center}
\includegraphics[angle=0,width=12cm]{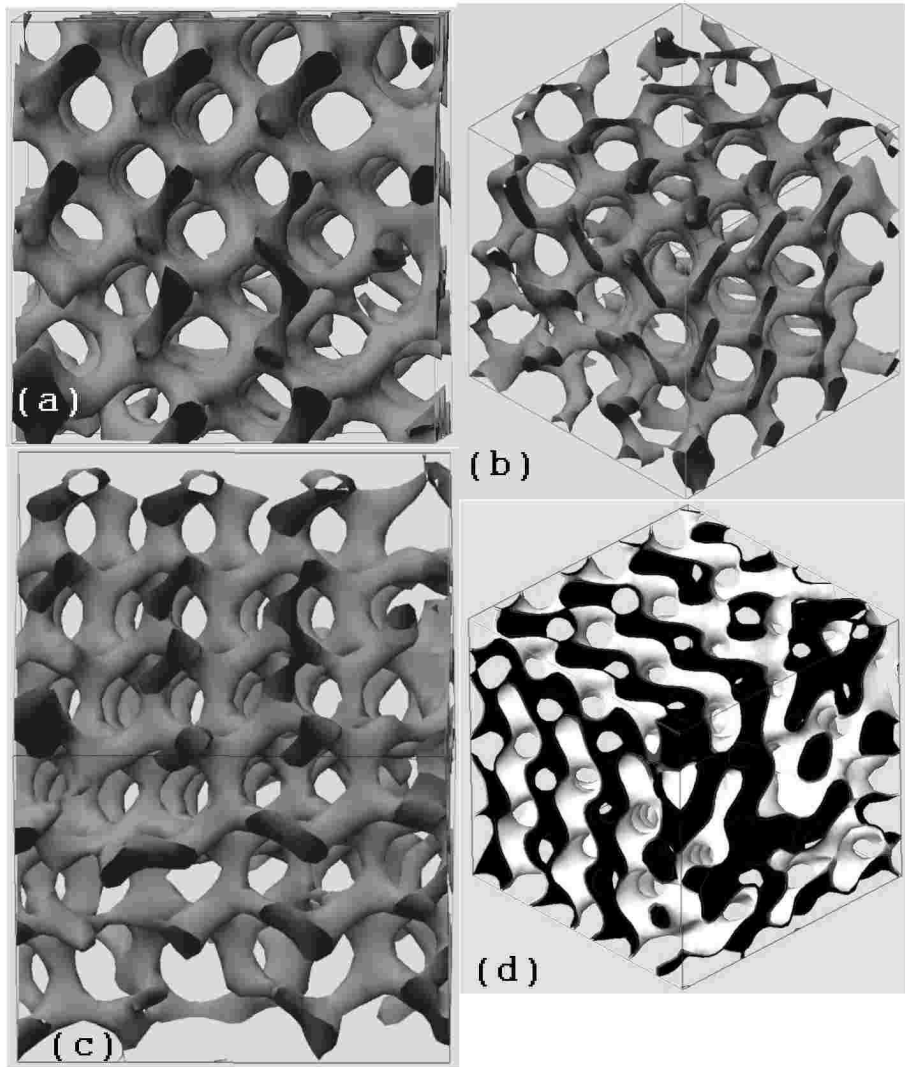}
\caption{Isosurfaces of the order parameter at time step $t=15000$ in
a highly ordered $33^3$ subdomain on a $128^3$ lattice for fluid
composition 09. Panels (a), (b) and (c) display the
$\phi(\mathbf{x})=0.40$ (in lattice units) minority phase isosurface
viewed as axonometric projections along the $(1\,0\,0)$,
$(1\,\overline{1}\,\overline{1})$ and $(1\,1\,0)$ directions,
respectively. Panel (d) shows the interface of the same lattice
subdomain along direction $(1\,\overline{1}\,\overline{1})$, where
black and white have been used to distinguish one immiscible fluid
phase from the other.} 
\label{GYROID_128}
\end{center}
\end{figure}

Figure \ref{GYROID_128} displays three viewpoints of the isosurface
$\phi=0.40$ for the same composition of fluid mixture~09 as simulated
on a $128^3$ lattice. We show the restriction to a $33^3$ subdomain,
all at time step $t=15000$, together with the oil-water
interface. Whereas on $64^3$ (or smaller) lattices the liquid
crystalline structure uniformly pervades the simulation cell, on
$128^3$ (and larger) lattices there are some imperfections present
resulting in ordered subdomains with slightly varying orientations
between which there exist domain boundaries. These boundaries can be
considered as defects in the structure, the presence of which is a
characteristic feature of liquid crystals.  A time scale for the
dynamics of some of these defects for our simulated gyroids
(simulation run 09) can be roughly estimated from
Fig.~\ref{SUR_COLOUR}: we observe for that particular slice that the  
topological genus of the interface changes in an interval ranging
between 500 and 1000 time steps. State-of-the-art visualisation proved
key in the analysis of results, and virtual reality technologies
can enhance its usefulness by increasing interactivity with the data
\cite{ReG}. 

\begin{figure}[htb]
\begin{center}
\begin{tabular}{r l r l}
(a)&\includegraphics[angle=0,width=6cm]{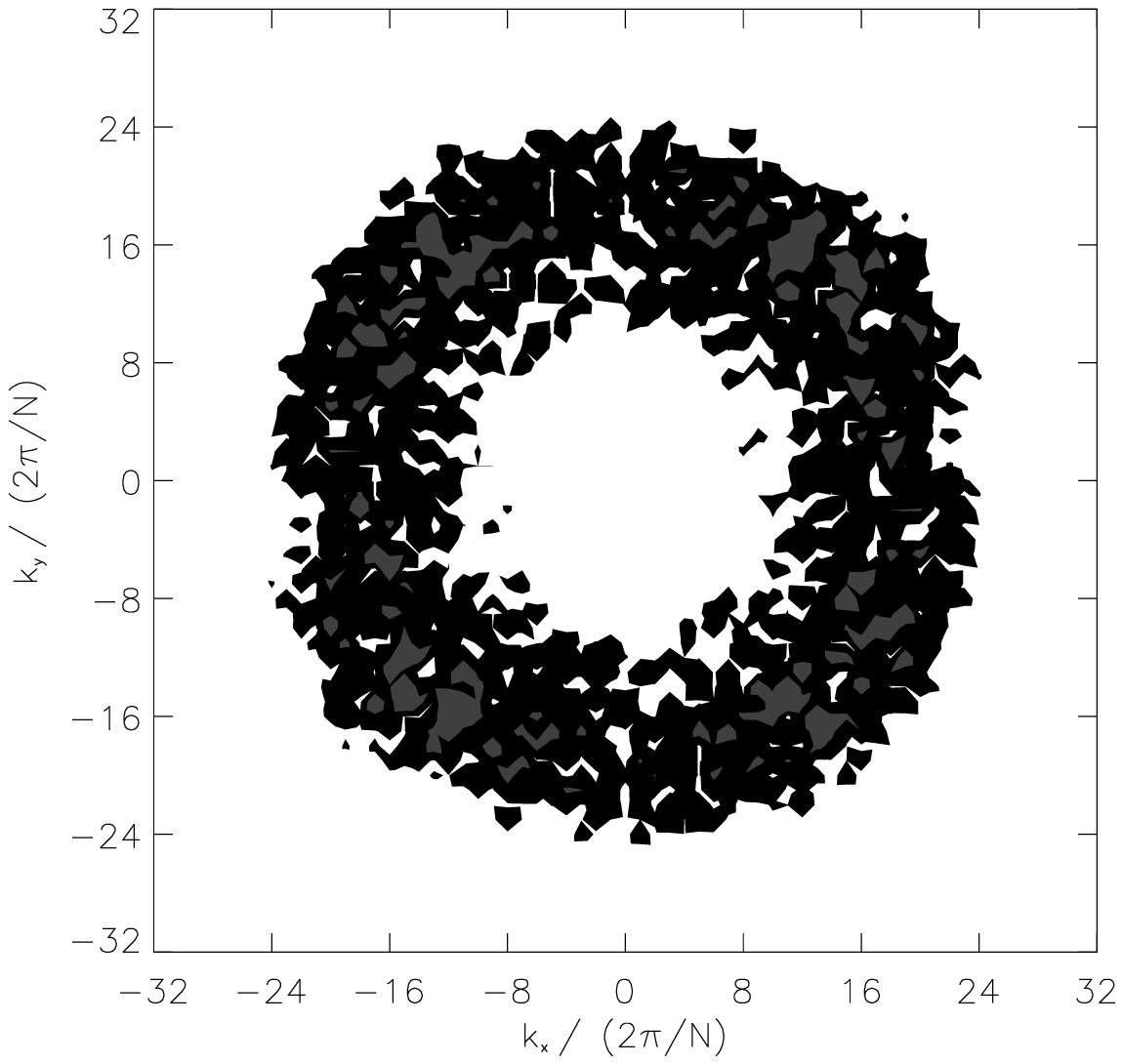}&
(b)&\includegraphics[angle=0,width=6cm]{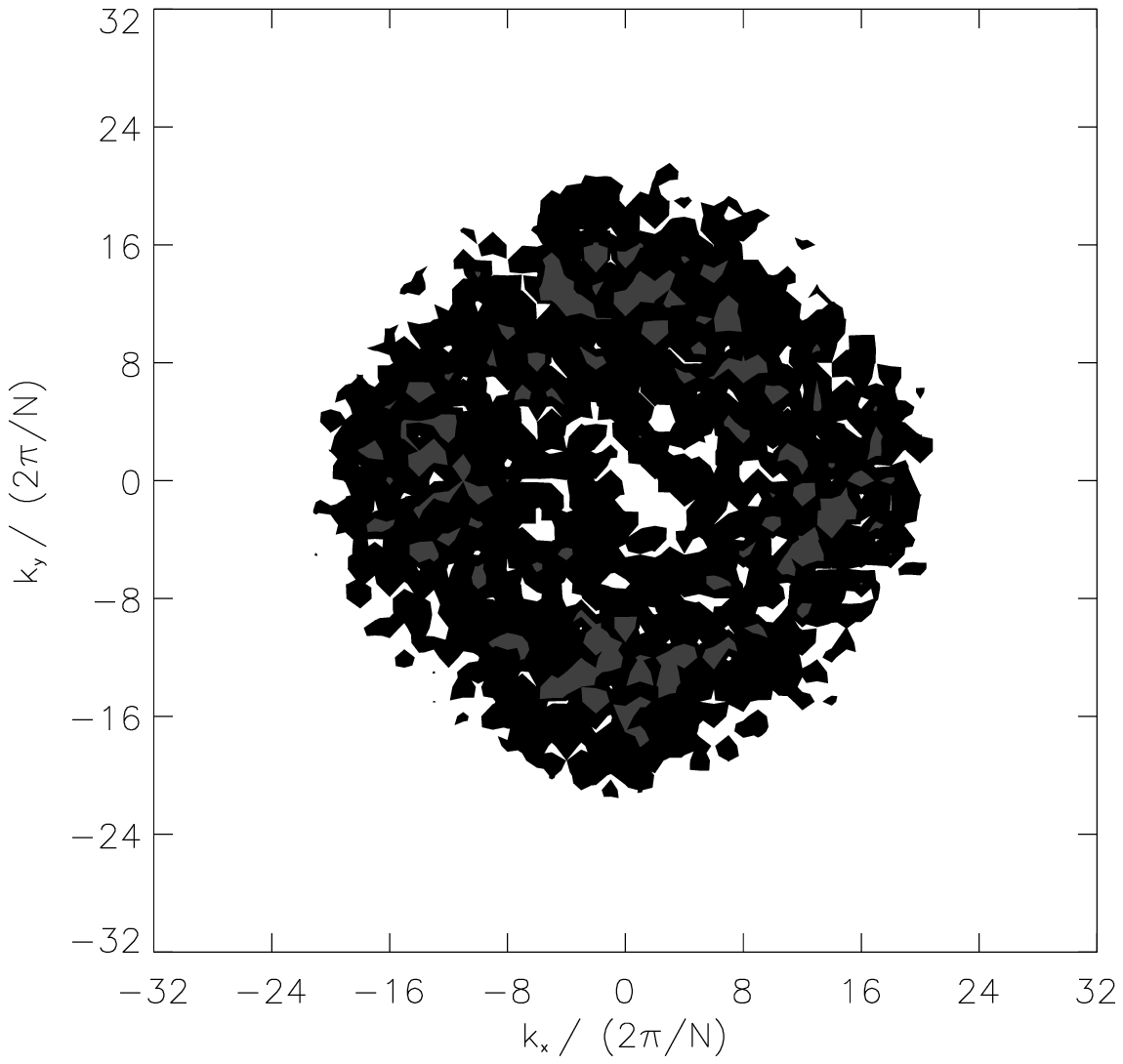}
\\
(c)&\includegraphics[angle=0,width=6cm]{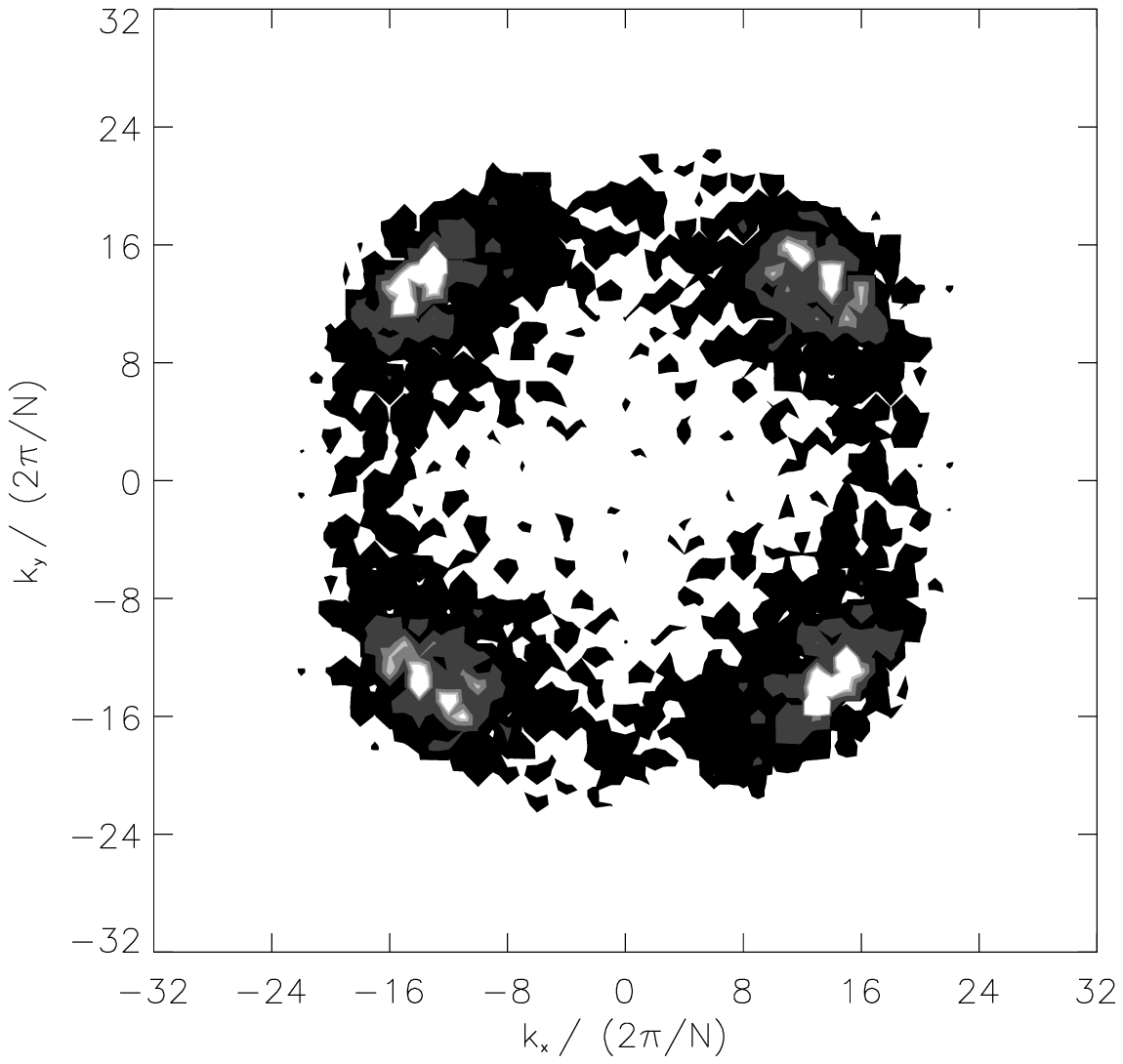}&
(d)&\includegraphics[angle=0,width=6cm]{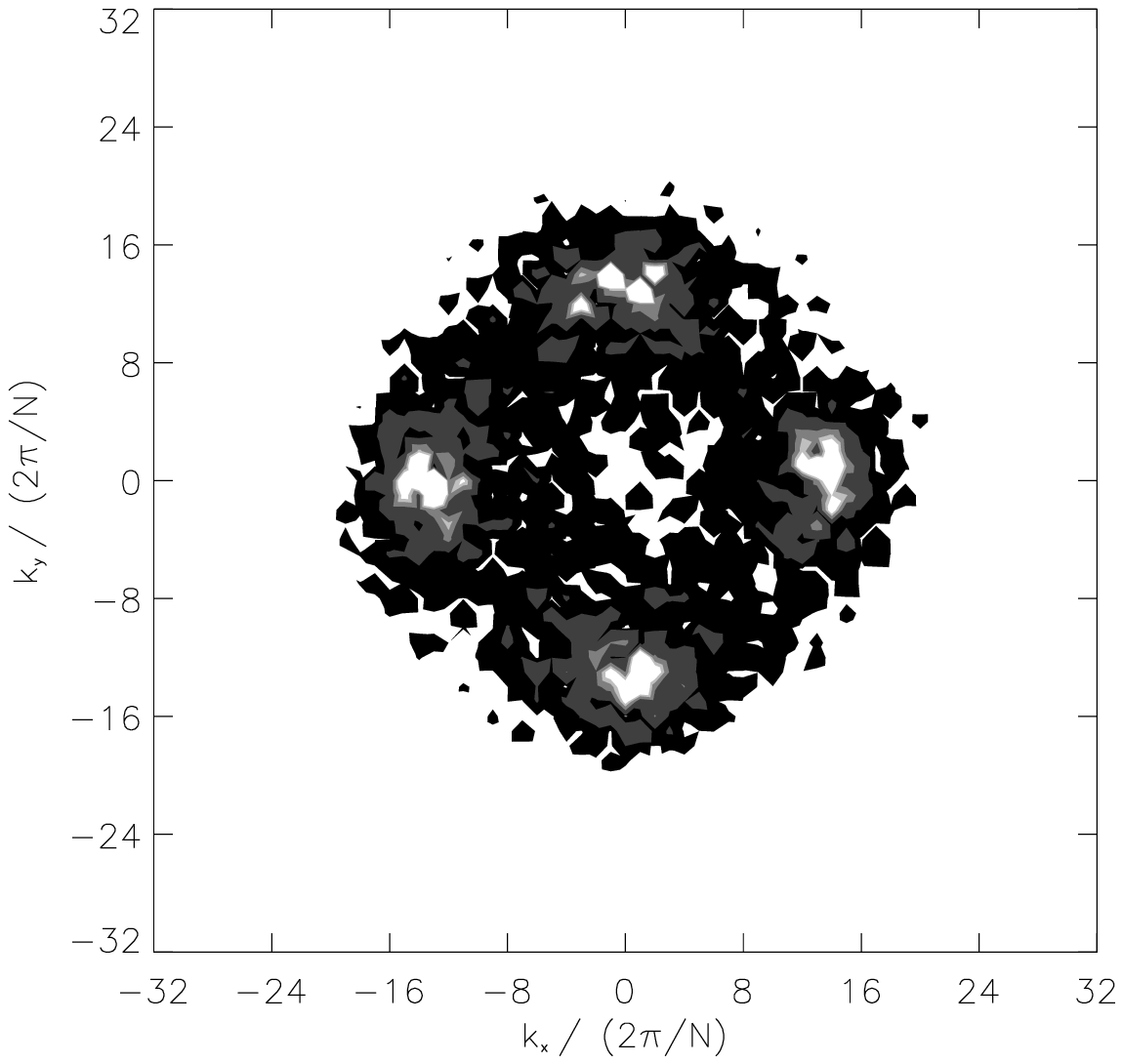}
\end{tabular}
\caption{Temporal evolution of $(k_x,k_y)$ slices of the structure
function $S(\mathbf{k})$ for fluid composition 09, viewed along the
$(\overline{1}\,0\,0)$ direction. Panels (a) and (b) show slices at
time step $t=500$, where $k_z/(2\pi/N)=0$ for (a) and
$k_z/(2\pi/N)=\pm14$ for (b). Similarly, panels (c) and (d) are slices
at time step $t=15000$ for $k_z/(2\pi/N)=0,\pm14$,
respectively; $N=128$ is the lateral lattice size. Shading denote
intensities $S=1,50$ and 100, where lighter greys up to white mean
higher intensities. The spherical shell structure in (a) and (b)
indicates the presence of a sponge (microemulsion) phase, which
becomes anisotropic at later times, (c) and (d). The superposition of
slices (c) and (d), namely, the ratio of peaks' positions of 
$\sum_{k_{\mathrm{z}}}S(\mathbf{k})$, are in full agreement with SAXS
experimental data for the gyroid mesophase. All quantities are in
lattice units.}        
\label{SF_SLICES}
\end{center}
\end{figure}

Small-angle x-ray scattering (SAXS) techniques have been widely used
in the determination of the nanostructure of fluid mesophases
\cite{LAURER,HAJDUK,SEDDON,KLINOWSKY}. SAXS spectra, or their
numerically computed versions \cite{GARSTECKI}, give peak patterns for
these mesophases that are used as fingerprints in determining unknown
structures. However, the lattice resolution of our simulations is
insufficient to detect multiple peak fingerprints in plots of the
spherically averaged structure function. Instead, its unaveraged
counterpart, $S(\mathbf{x})$, shows complete agreement of ratios of
reciprocal vector moduli with those observed in diffraction patterns
of the gyroid, as we display in Fig.~\ref{SF_SLICES} for fluid
composition 09 \cite{HAJDUK}. In addition, visual inspection of the
unit cell of the oil-water interface unequivocally identifies it with
that of the gyroid. The size of such a unit cell as seen in optical
textures allows us to associate a length scale to the lattice for a
particular experimental realisation. For the system reported by Hajduk
{\em et~al.}~\cite{HAJDUK}, the lattice would need to be 291~nm in
side length with a resolution of 2.3 nm per lattice unit. 

Although previous simulation papers on amphiphilic mixtures using
free-energy based Langevin diffusion equations have reported the
reproduction of structures resembling the gyroid
\cite{VLIMMEREN,ZVELINDOVSKY}, none of them have studied its features
or dynamics, or incontestably demonstrated its gyroid morphology. 
Furthermore, in one of these articles we observe that the fluid 
mesostructure is not stationary \cite{ZVELINDOVSKY}, whereas by
eyeballing the whole simulation cell in another \cite{VLIMMEREN} one
becomes aware that the structure has a morphology which is reminiscent
of the molten gyroid we describe here.

In our simulations we observe that, at late times, the gyroid is
much closer to stationarity than the sponge mesophase: for equal time
slices in their evolution, temporal changes of the mesostructure over
a period of 1000 time steps are considerable for the sponge
(e.g. simulation run 06) whereas for the gyroid (e.g. simulation run
09) they appear as slight interfacial rearrangements and undulations,
reminiscent of breathing modes, keeping the variation in the
position of each unit cell small compared to the lattice size
$N$. This late-time (ca. 30000 time steps) structural dynamics is
characterised in our simulations by the fact that the topological
genus is (statistically) preserved for the gyroid; for the sponge, it
is not. This can be understood as structural stabilisation by rigidity
in the gyroid, and a flowing, glassy dynamics for the sponge. Such a
distinctive behaviour for the sponge may have a bearing 
on its density fluctuations and render them different to those
occurring in a topology-conserving dynamics. It is therefore not
surprising that (a) we found the oscillation modes for the sponge
(simulation run 06) to be at least one order of magnitude more intense
than those for the different gyroids we simulated (simulation runs 08,
09 and 10, {\em cf.} Fig.~\ref{SFFT_09.12.15.16_ALL-FREQS}), and (b)
recent experimental studies, using dynamic light scattering and
various relaxation methods \cite{SCHWARZ}, do not report on
fluctuations for the gyroid \cite{SQUIRES}, whereas they do for the
sponge mesophase. 

It is accepted wisdom \cite{SEUL}, and a working hypothesis in many
simulation studies \cite{SAGUI,ROLAND,NONOMURA}, that periodically
modulated phases may arise in fluid mixtures whenever a repulsive
long-range interaction competes with an attractive short-range one for
a configuration that minimises the  interfacial Hamiltonian, possibly
also in the presence of a thermal, entropic contribution. Little is
said about whether non-locality is not only a sufficient but also a
necessary condition, or whether the non-locality of the relevant model
needs to be imposed {\em ab initio} or is rather an effective emergent
feature picked up by the order-parameter autocorrelation function. The
LB model we employ in this paper only incorporates local 
interactions in its mesodynamics; this feature allows its algorithm
to be easily parallelised and achieve exceptionally good performance
\cite{COMPARISON}. Nonetheless, we have demonstrated that the model is
able to simulate liquid crystalline, cubic mesophases, such 
as the gyroid in binary immiscible fluid mixtures with an amphiphile,
and the primitive ``P" in binary, amphiphile-solvent mixtures
\cite{JACS_P}, whose density-density correlations are markedly
non-local, and in the formation of which hydrodynamic interactions
play a vital role. Non-locality, in our case, is an emergent property
of a local model.

It is worthwhile pointing out that Prinsen {\em et al.}
\cite{PRINSEN}, using a DPD model and basing their claims on Monte
Carlo studies of equilibrium cubic phases by Larson \cite{LARSON},
suggested that cubic phases could be engineered to appear if their
bead-rod model were elaborated beyond dimers. If fact, Groot \&
Madden's (inconclusive) finding of a gyroid-like structure with a DPD
model for bead-spring chain copolymers \cite{GROOT2} might be
considered to support such an assertion. Part of the importance of our
simulations, as well as those of Nekovee \& Coveney \cite{JACS_P}, is
to refute such conjectures, by demonstrating that cubic mesophases
arise in very simple, minimalist, athermal and
hydrodynamically-correct fluid models, with a locally-interacting
vector order parameter and reproducing universal behaviour.

\section{Summary and conclusions}
\label{CONCLU}

Our simulations furnish the first quantitative account of phase
segregation kinetics and mesophase self-assembly in amphiphilic fluids
with a three-dimensional model based on the Boltzmann transport
equation. The method is hydrodynamically correct, athermal, and models
the amphiphilic species as bipolar, point-like particles experiencing
short-range interactions with mean fields created by the surrounding
binary immiscible (``oil-water") and amphiphilic medium. 

We studied the phase segregation pathway in a homogeneous
oil-water-surfactant mixture at composition $1:1:x$, respectively,
where $0\le x\le1.3$ is the surfactant-to-water (or to oil) mass
fraction. We observed segregation slowdown in the average size of
oil-water domains with increasing $x$, and the reproduction of known
crossovers, namely, from algebraic to logarithmic to stretched
exponential functions. This confirms the usefulness of our method in
apprehending the fundamental phenomenology of amphiphilic fluid
mixtures; the presence of transients in these crossovers is gratifying
given their experimental observation. In order to rule out an increase
in total density as $x$ is increased as a factor contributing to the
slowdown along with the reduction in surface tension, future work
should investigate domain growth at constant total density. 

The stretched exponential functional form occurring at domain growth 
arrested regimes can be ascribed to the accumulation of a large number
of relaxation modes associated with surfactant dynamics onto and at
the oil-water interfaces \cite{WEIG97}.  The late-time structure at
these regimes are (a) disordered non-stationary sponge mesophases if
the initial amphiphile concentration is lower than a threshold region
or (b) well-defined liquid-crystalline cubic gyroid mesophases of
pinned  domain sizes with defects if this initial concentration is
higher than such a threshold. We also found thresholds in the
surfactant coupling strengths, $|g_{ss}|$ and $|g_{bs}|$, for a
sponge-to-gyroid  transition. In the transition region we observed a
crossover structure sharing the structural features of both the gyroid
and sponge. We also found that, for the number of time steps
simulated, both sponge and gyroid exhibit undamped oscillations at all
length scales. For some length scales, the temporal trend of their
Fourier amplitudes is to slowly die out; for others, it increases.
 
We found that extremely slow domain growth can be mistaken for genuine
arrested growth if attention is not paid to minute length
scales. Truly segregation-halted regimes exhibit oscillations in
average domain size, which can be seen at sufficiently late
times. These oscillations are caused partly by Marangoni flows
generated by inhomogeneities in the surfactant adsorbed on the
oil-water interface, and partly by a surfactant dynamics dictated by
competing mechanisms,  namely, surfactant attraction towards the
interface and surfactant-surfactant interactions. Because our model
does not presuppose that all the surfactant is adsorbed on the
interface, as Langevin approaches based on the adiabatic
approximation do \cite{GOMPPER,NONOMURA,SEUL}, surfactant-surfactant
interactions are not limited to repulsion. Hence, in regimes of large
surfactant concentration, and especially in those for which regions of
the (oil-water) interface can 
be sufficiently close to each other, surfactant is not constrained to
dwell on the interface; rather, it is reasonable to propose the
existence of an adsorption-desorption dynamics driving surfactant
towards and away from it. Our results showing that (a) an increase in
$|g_\mathrm{ss}|$ excites higher frequencies, (b) an increase in
$|g_\mathrm{bs}|$ dampens most frequencies, and (c) there appear
surfactant currents bridging adjacent interfacial regions, confirm
this proposal.

Our method is not only the first lattice-Boltzmann model to deal with
segregation kinetics in three-dimensional amphiphilic fluid mixtures,
but the first complex fluid model to unequivocally reproduce the
gyroid cubic mesophase, using a high level of abstraction in modelling
the amphiphile. The truly mesoscopic, particulate nature of the  
surfactant in this model accounts for the complex, dynamical behaviour
observed, even in a noiseless scenario like ours. It is not surprising
that Ginzburg-Landau-based Langevin models which treat surfactant
implicitly through a scalar parameter modifying the free energy 
are only able to exhibit oscillations which decay rapidly in
time and whose frequency spectrum has but one peak. Our simulations of
liquid crystalline mesophases prove that, contrary to what has been
previously claimed \cite{SAGUI,ROLAND,SEUL}, surfactant-surfactant
interactions need not be long ranged in order for periodically
modulated, long-range ordered structures to self-assemble. In
addition, our findings rebut the suggestion \cite{PRINSEN} that cubic
mesophases can be simulated only when the amphiphile is modelled with
a high degree of molecular specificity.

The simulation of the sponge and gyroid phases, and their complex
oscillatory dynamics, confirms the richness of our model's parameter
space. Our lattice-Boltzmann model provides a kinetically and
hydrodynamically correct, bottom-up, mesoscale description of the
generic behaviour of amphiphilic fluids, which is also extremely
computationally efficient on massively parallel platforms. Future
extensions of this work include the search for regimes leading to
equilibrium mesophases of more varied symmetries, the study of
shear-induced symmetry transitions, and large scale studies of defect
dynamics in liquid crystalline phases. In fact, the TeraGyroid
project, a successful Grid-based transatlantic endeavour employing  
more than 6000 processors and 17 teraflops at six supercomputing
facilities \cite{FLYER}, has its scientific {\em raison d'\^{e}tre}
based on the results we report in this paper. TeraGyroid proves the
value of computational steering tools \cite{COMPUSTEERING} in mapping
new parameter space regions of our model.

\section{Acknowledgments}
This work was supported by the UK EPSRC under grants GR/M56234 and 
RealityGrid GR/R67699 which provided access to SGI Origin2000 and 
SGI Origin3800 supercomputers at Computer Services for Academic 
Research (CSAR), Manchester Computing Consortium, UK. We also thank 
the Higher Education Funding Council for England (HEFCE) for our 
on-site 16-node SGI Onyx2 graphical supercomputer. We thank
Dr~Rafael Delgado Buscalioni for enlightening discussions and 
Dr~Keir Novik for technical assistance. NGS also wishes to thank 
Prof~David Jou, Prof~Jos\'e Casas-V\'azquez and Dr~Juan Camacho at 
the Universitat Aut\`onoma de Barcelona, Spain, for their support.

\newpage

\newpage

\newpage

\newpage

\newpage

\newpage

\newpage

\newpage

\newpage

\newpage

\newpage

\newpage

\newpage

\newpage


\begin{thebibliography}{99}

\bibitem{GELBART}
W. M. Gelbart, D. Roux, and A. Ben-Shaul, {\em Modern ideas and
problems in amphiphilic science.} (Springer, Berlin, 1993).

\bibitem{AMPHI_BIO} 
D. Chapman and M. N. Jones, {\em Micelles, Monolayers, and
Biomembranes.} (Wiley, John \& Sons, 1994).

\bibitem{MICROEMUL_INDUST}
P. Kumar and K. L. Mittal (eds), {\em Handbook of Microemulsion
Science and Technology.} (Marcel Dekker, New York, 1999).

\bibitem{GOMPPER&SCHICK2}
G. Gompper and M. Schick, ``Microscopic models of microemulsions.'' 
In {\em Handbook of microemulsion science and technology.}
P. Kumar \& K. L. Mittal (eds) (Marcel Dekker, New York, 1999.) 

\bibitem{MARIANI} P. Mariani, V. Luzzati and H. Delacroix,
J. Mol. Biol. {\bf 204}, 165 (1988).

\bibitem{MARRINK} S.-J. Marrink and D. P. Tieleman, J. Am. Chem. Soc. 
{\bf 123}, 12383 (2001).

\bibitem{LUZZATI} V. Luzzati, R. Vargas, P. Mariani, A. Gulik, 
and H. Delacroix, J. Mol. Biol {\bf 229}, 540 (1993).

\bibitem{GOMPPER&SCHICK3} G. Gompper and M. Schick, ``Self-assembling
amphiphilic systems.'' In {\em Phase Transitions and Critical
Phenomena.} C. Domb and J. Lebowitz (eds), Vol. 16, pages 1-176
(Academic Press, London, 1994).

\bibitem{P2} N. Gonz\'alez-Segredo, M. Nekovee, and P. V. Coveney, 
Phys. Rev. E {\bf 67}, 046304 (2003).

\bibitem{BRAY} A. J. Bray, Adv. Phys. {\bf 43}, 357-459 (1994).

\bibitem{KAWAKATSU93} T. Kawakatsu, K. Kawasaki, M. Furusaka,
H. Okabayashi, and T. Kanaya, J. Chem. Phys. {\bf 99}, 8200 (1993).  

\bibitem{LARADJI92} M. Laradji, H. Guo, M. Grant, and
M. Zuckermann, J. Phys. A {\bf 24}, L629 (1991); M. Laradji, H. Guo, 
M. Grant, and M. Zuckermann, J. Phys.: Cond. Matt. {\bf 4}, 6715 
(1992).  

\bibitem{EMERTON97} A. N. Emerton, P. V. Coveney, and B. M. Boghosian,
Phys. Rev. E {\bf 55}, 708 (1997).

\bibitem{SEDDON} J. M. Seddon and R. H. Templer, in {\em Handbook
of Biological Physics.} R. Lipowsky and E. Sackmann (eds)
(Elsevier Science B. V., London, 1995.) Vol. 1, pp97-153.

\bibitem{HOHENBERG} P. C. Hohenberg and B. I. Halperin,
  Rev. Mod. Phys. {\bf 49}, 435 (1977).
 

\bibitem{HALPERIN} B. I. Halperin, P. C. Hohenberg, and S. K. Ma,
  Phys. Rev. B. {\bf 10}, 139 (1974).
 

\bibitem{YAO} J. H. Yao and M. Laradji, Phys. Rev. E {\bf 47}, 2695
(1993). 

\bibitem{LARADJI94} M. Laradji, O. G. Mouritsen, S. Toxvaerd, and
J. Zuckermann, Phys. Rev. E {\bf 50}, 1243 (1994). 

\bibitem{WEIG97} F. W. J. Weig, P. V. Coveney, and B. M. Boghosian,
Phys. Rev. E {\bf 56}, 6877 (1997). 

\bibitem{LOVE01} P. J. Love, P. V. Coveney, and B. M. Boghosian,
Phys. Rev. E {\bf 64}, 021503 (2001). 

\bibitem{GROOT2} R. D. Groot and T. J. Madden, J. Chem. Phys. {\bf 108}, 8713 (1998).

\bibitem{GROOT} R. D. Groot, T. J. Madden, and D. J. Tildesley,
J. Chem. Phys. {\bf 110}, 9739 (1999).

\bibitem{PRINSEN} P. Prinsen, P. B. Warren, and M. A. J. Michels,
Phys. Rev. Lett. {\bf 89}, 148302 (2002).

\bibitem{JACS_P} M. Nekovee and P. V. Coveney, J. Am. Chem. Soc. {\bf 123},
12380 (2001).

\bibitem{NONOMURA} M. Nonomura and T. Ohta, J. Phys.:
Condens. Matt. {\bf 13}, 9089 (2001).

\bibitem{IMAI} M. Imai, A. Saeki, T. Teramoto, A. Kawaguchi,
K. Nakaya, T. Kato, and K. Ito, J. Chem. Phys. {\bf 115}, 10525
(2001). 

\bibitem{QI} S. Qi and Z.-G. Wang, Phys. Rev. E. {\bf 55}, 1682
(1997). 

\bibitem{GOMPPER} G. Gompper and M. Hennes, J. Phys. II France {\bf
4}, 1375 (1994).

\bibitem{GOMPPER2} G. Gompper and J. Goos, Phys. Rev. E. {\bf
50}, 1325 (1994).

\bibitem{OHTA} T. Ohta and K. Kawasaki, Macromolecules {\bf 19}, 
2621 (1986).

\bibitem{ZVELINDOVSKY} A. V. Zvelindovsky, G. J. A. Sevink, and
J. G. E. M. Fraaije, Phys. Rev. E. {\bf 62}, R3063 (2000).

\bibitem{VLIMMEREN} B. A. C. van Vlimmeren, N. M. Maurits,
A. V. Zvelindovsky, G. J. A. Sevink, and J. G. E. M. Fraaije,
Macromolecules {\bf 32}, 646 (1999).

\bibitem{COVNOBEL} P. V. Coveney, Phil. Trans. R. Soc. Lond. A {\bf 361},
1057 (2003).

\bibitem{NADIM} D. Gueyffier, J. Lie, A. Nadim, R. Scardovelli, and
  S. Zaleski, J. Comput. Phys. {\bf 152}, 423 (1999).

\bibitem{SUCCI} S. Succi, {\em The lattice-Boltzmann equation--for
  fluid dynamics and beyond.} (Oxford University Press, Oxford, 2001.)

\bibitem{CHEN_AMPHI} H. Chen, B. M. Boghosian, P. V. Coveney, and
M. Nekovee, Proc. Roy. Soc. Lond. A {\bf 456}, 2043 (2000). 

\bibitem{MAZIAR-ET-AL} M. Nekovee, P. V. Coveney, H. Chen, and
B. M. Boghosian, Phys. Rev. E {\bf 62}, 8282 (2000). 

\bibitem{LAMURA} A. Lamura and G. Gonnella, Int. J. Mod. Phys. C 
{\bf 22}, 1469 (1998); A. Lamura, G. Gonnella, and J. M. Yeomans,
Europhys. Lett. {\bf 45}, 314 (1999).

\bibitem{THEISSEN} O. Theissen, G. Gompper, and D. M. Kroll,
Europhys. Lett. {\bf 22}, 419 (1998); O. Theissen and G. Gompper, 
Eur. Physical J. B {\bf 43}, 91 (1999).

\bibitem{COMPARISON} The scaling with the number of processing
  elements (PE) for our LB model's implementation  is close to
  linearity and superlinear on more than 64 PEs on Cray T3E-1200E and
  SGI Origin 3800 supercomputers, respectively; {\em cf.} P.~J.~Love,
  M. Nekovee, P. V. Coveney, J. Chin, N. Gonz\'alez-Segredo, and
  J. M. R. Martin, Comp. Phys. Commun. {\bf 153}, 340-358 (2003).

\bibitem{ENTROPIC} No entropic LB scheme for immiscible fluid mixtures
has been reported to date. See 
I. V. Karlin, A. Ferrante, and H. C. \"Ottinger, Europhys. Lett. {\bf
47}, 182 (1999); H. Chen and C. Teixeira, Comp. Phys. Comm. {\bf 129},
21 (2000); B. M. Boghosian, J. Yepez, P. V. Coveney, and A. Wagner, 
Proc. Roy. Soc. Lond. A {\bf 457}, 717 (2001); B. M. Boghosian,
P. J. Love, P. V. Coveney, I. V. Karlin, S. Succi, and J. Yepez,
Phys. Rev. E {\bf 68}, 025103(R) (2003).

\bibitem{FRISCH} U. Frisch, D. d'Humi\`eres, B. Hasslacher,
P. Lallemand, Y. Pomeau, and J.-P. Rivet, Complex Systems {\bf 1}, 649
 (1987). 

\bibitem{SHAN-CHEN} X. Shan and H. Chen, Phys. Rev. E {\bf 47}, 1815
(1993); X. Shan and H. Chen, Phys. Rev. E {\bf 49}, 2941 (1994).  

\bibitem{MPI} URL: {\tt http://www.mpi.org/}~.

\bibitem{IAIN} I. Murray, {\em Lattice Boltzmann study of amphiphilic
fluids.} Centre for Computational Science Internal Report, Department
of Chemistry, Queen Mary, University of London, October 2001.  

\bibitem{ROWLINSON} J. S. Rowlinson and B. Widom, {\em The molecular
  theory of capillarity.} (Clarendon Press, Oxford, 1982.)

\bibitem{FERZIGER} J. H. Ferziger and H. G. Kaper, {\em Mathematical 
theory of transport processes in gases} (North Holland, Amsterdam, 
1972). 

\bibitem{ALLEN-TILDESLEY} M. P. Allen and D. J. Tildesley, {\em
  Computer simulations of liquids.} (Clarendon Press, Oxford, 1990.)

\bibitem{GUNTON} J.~D. Gunton, M. San Miguel, and P.~S. Sahni, in {\em
Phase Transitions and Critical Phenomena.} C. Domb \& J. Lebowitz
(eds), Vol.~8. (Academic Press, New York, 1983).

\bibitem{GOMPPER&SCHICK} G. Gompper, ``Bulk and interfacial
properties of amphiphilic systems: A Ginzburg-Landau approach." In
{\em Structure and dynamics of strongly interacting colloids and
supramolecular aggregates in solution.} S.-H. Chen, J. S. Huang, 
\& P. Tartaglia (eds) (Kluwer Academic Publishers, Dordrecht, 
1992).
   

\bibitem{KOVALCHUK} V. I. Kovalchuk, H. Kamusewitz, D. Vollhardt, and 
N. M. Kovalchuk, Phys. Rev. E {\bf 60}, 2029 (1999).

\bibitem{LAURER} J. H. Laurer, D. A. Hajduk, J. C. Fung, J. W, Sedat,
S. D. Smith, S. M. Gruner, D. A. Agard, and R. J. Spontak,
Macromolecules {\bf 30}, 3938 (1997).

\bibitem{COMPUSTEERING} J. Chin, J. Harting, S. Jha, P. V. Coveney,
     A. R. Porter, and S. M. Pickles, Contemp.~Phys. {\bf 44}, 417
     (2003).

\bibitem{ReG} URL: {\tt http://www.RealityGrid.org}~.

\bibitem{HAJDUK} D. A. Hajduk, P. E. Harper, S. M. Gruner, 
C. C. Honeker, G. Kim, and E. L. Thomas,
Macromolecules {\bf 27}, 4063 (1994).

\bibitem{KLINOWSKY} J. Klinowsky, A. L. MacKay, and H. Terrones,
Phil. Trans. R. Soc. Lond. A {\bf 354}, 1975 (1996).

\bibitem{GARSTECKI} P. Garstecki and R. Ho\l{}yst, J. Chem. Phys.
{\bf 115}, 1095 (2001).

\bibitem{SCHWARZ} B. Schwarz, G. M\"onch, G. Ilgenfritz, and R. Strey,
Langmuir {\bf 16}, 8643 (2000).

\bibitem{SQUIRES} A. M. Squires, R. H. Templer, J. M. Seddon, 
J. Woenckhaus, R. Winter, S. Finet, and N. Theyencheri, 
Langmuir {\bf 18}, 7384 (2002).

\bibitem{SEUL} M. Seul and D. Andelman, Science {\bf 267}, 476 (1995).

\bibitem{SAGUI} C. Sagui and R. C. Desai, Phys. Rev. Lett. {\bf 71}, 
3995 (1993); C. Sagui and R. C. Desai, Phys. Rev. E {\bf 49}, 2225 (1994);
C. Sagui and R. C. Desai, Phys. Rev. E {\bf 52}, 2807 (1995).

\bibitem{ROLAND} C. Roland and R. C. Desai, Phys. Rev. B {\bf 42}, 6658 (1990).

\bibitem{LARSON} R. G. Larson, J. Phys. II France {\bf 6}, 1441 (1996).

\bibitem{FLYER} See, e.g., {\em TeraGyroid:~Grid-based
  lattice-Boltzmann simulations of defect dynamics in amphiphilic
  liquid crystals}, demonstration at the SuperComputing2003
  conference, Phoenix (Arizona, USA), 15th--21st November 2003,  
  URLs:~{\tt
http://www.sc-conference.org/sc2003/~, http://www.RealityGrid.org/news.html~}. 




\end{thebibliography}
\end{document}